%% file: ms.tex
\pdfoutput=1
\RequirePackage{fix-cm}
\documentclass[smallextended]{svjour3}       
\smartqed  
\usepackage{graphicx}
\usepackage{balance}
\usepackage{framed}
\usepackage{epsfig}
\usepackage{color}
\usepackage{subfigure}
\usepackage{multirow,tabularx}
\usepackage{amssymb}
\usepackage{mathtools}
\usepackage{graphicx}
\usepackage{epstopdf}
\usepackage{cite}
\usepackage{multirow}
\usepackage{bm}
\usepackage{hhline}
\usepackage{csquotes}
\usepackage{array}
\usepackage[yyyymmdd,hhmmss]{datetime}
\usepackage[subject={Todo}]{pdfcomment}
\usepackage[textsize=footnotesize,bordercolor=black!20]{todonotes}
\usepackage{xcolor,colortbl}
\usepackage{amssymb}
\usepackage{pifont}
\usepackage[algo2e,titlenumbered,ruled]{algorithm2e} 
\usepackage{lipsum,environ,amsmath}
\usepackage{slashbox,amsmath}
\usepackage{xspace}
\usepackage[round, sort, numbers, authoryear]{natbib}
\usepackage{hyperref}

\usepackage{bbm}

\newenvironment{problem1}[1][htb]
  {
  \begin{algorithm2e}[#1]%
  }{\end{algorithm2e}}

\newcommand{\numsquishlist}{
   \begin{list}{\arabic{Lcount}. }
    { \usecounter{Lcount}
 \setlength{\itemsep}{-.1ex}      \setlength{\parsep}{0ex}
      \setlength{\topsep}{0ex}       \setlength{\partopsep}{0ex}
      \setlength{\leftmargin}{1em} \setlength{\labelwidth}{1em}
      \setlength{\labelsep}{0.1em} } }
\newcommand{\numsquishend}{\end{list}}

\newcommand{\squishlist}{
   \begin{list}{$\bullet$}
    { \setlength{\itemsep}{-.1ex}      \setlength{\parsep}{0ex}
      \setlength{\topsep}{0ex}       \setlength{\partopsep}{0ex}
      \setlength{\leftmargin}{.8em} \setlength{\labelwidth}{1em}
      \setlength{\labelsep}{0.5em} } }
\newcommand{\squishend}{\end{list}}
\definecolor{Gray}{gray}{0.85}
\definecolor{LightGray}{rgb}{0.9,0.9,0.9}
\definecolor{LightBlue}{rgb}{0.8,0.8,1}

\newcommand{\slip}{{\sc Mining Patterns of Leadership Dynamics\xspace}}%
\newcommand{\sfip}{{\sc Mining Patterns of Followership Dynamics\xspace}}%

\usepackage{setspace}

 \journalname{Social Network Analysis and Mining}

\def\makeheadbox{{%
\hbox to0pt{\vbox{\baselineskip=10dd\hrule\hbox
to\hsize{\vrule\kern3pt\vbox{\kern3pt
\hbox{\bfseries [Social Network Analysis and Mining]}
\hbox{This is a post-peer-review, pre-copyedit version of this article.}
\hbox{The final authenticated version is available online at: \href{https://doi.org/10.1007/s13278-019-0600-z}{https://doi.org/10.1007/s13278-019-0600-z}.}
\kern3pt}\hfil\kern3pt\vrule}\hrule}%
\hss}}}

\begin{document}

\title{Mining and Modeling Complex Leadership-Followership Dynamics of Movement data
}

\titlerunning{Mining and Modeling Complex Leadership-Followership Dynamics}        

\author{Chainarong Amornbunchornvej         \and       Tanya Y. Berger-Wolf 
}


\institute{C. Amornbunchornvej \at
              National Electronics and Computer Technology Center (NECTEC) \\
              Pathum Thani, Thailand.\\
              \email{chainarong.amo@nectec.or.th}           
           \and
           Tanya Y. Berger-Wolf \at
              Department of Computer Science\\
              University of Illinois at Chicago, Chicago, IL, USA.\\
              \email{tanyabw@uic.edu}
}

\date{Received: 30 November 2018 / Revised: 6 June 2019 / Accepted: 16 September 2019 / Published: 03 October 2019}

\maketitle

\begin{abstract}
  Leadership and followership are essential parts of collective decision and organization in social animals, including humans. In nature, relationships of leaders and followers are dynamic and vary with context or temporal factors. Understanding dynamics of leadership and followership, such as how leaders and followers change, emerge, or converge, allows scientists to gain more insight into group decision-making and collective behavior in general. However, given only data of individual activities, it is challenging to infer the dynamics of leaders and followers. In this paper, we focus on mining and modeling frequent patterns of leading and following. We formalize new computational problems and propose a framework that can be used to address several questions regarding group movement. We use the leadership inference framework, mFLICA, to infer the time series of leaders and their factions from movement datasets, then propose an approach to mine and model frequent patterns of both leadership and followership dynamics. We evaluate our framework performance by using several simulated datasets, as well as  the real-world dataset of  baboon movement to demonstrate the applications of our framework.  These are novel computational  problems and, to the best of our knowledge, there are no existing comparable methods to address them. Thus, we modify and extend an existing leadership inference framework to provide a non-trivial baseline for comparison. Our framework performs better than this baseline in all datasets. Our framework opens the opportunities for scientists to generate testable scientific hypotheses about the dynamics of leadership in movement data. 
\end{abstract}

 \input{02-intro}

\input{03-probstate}
 \input{04-method}
 \input{05-experiment}
 \input{06-results}
 \input{07-conclusion}

\balance
\bibliographystyle{spbasic}

\input{ms.bbl}\end{document}

%% file: 02-intro.tex
\section{Introduction}

Leadership is a process that leaders influence followers' actions in order to achieve the collective goal~\cite{hogg2001social,glowacki2015leadership}.  Leadership is an essential part that fosters success of coordinated behaviors in social species~\cite{glowacki2015leadership,couzin2005effective,hogg2001social}, such as foraging, migration, territorial defense, and so on. In most species, leadership is not permanent but may change with context (the one who leads the group to food or water may be different from the one who leads the flight from a predator) or other social circumstances (two rivaling subgroups may come to a joint decision and merge under single leadership or, vice versa, a group may split to explore several directions). 
Understanding dynamics of leadership, such as how leaders change, emerge, or converge, allows scientists to gain more insight into group decision-making and collective behavior in general. In this paper, we focus on mining and modeling frequent patterns of leadership dynamics.  

One of the intuitive definitions of leadership that is commonly found in nature is the initiation of coordinated activities~\cite{krause2000leadership,Smith2015187,stueckle2008follow}. In the context of movement, leaders are initiators who initiate coordinated movement that everyone follows~\cite{mFLICASDM18}.
There are several works have been developed to infer leaders from time series of movement data, such as FLOCK method~\cite{andersson2008reporting}, LPD framework~\cite{kjargaard2013time}, and methods based on a dynamic following network concept~\cite{Amornbunchornvej:2018:CED:3234931.3201406} and ~\cite{mFLICASDM18}. 

Nevertheless, the challenges in the field still remain regarding how to infer and model {\em the dynamics of the frequent patterns} of leadership events. For example, suppose $i$ and $j$ lead separate sub-groups, how often do the two groups merge to a larger group lead by $k$? How likely is it that the group lead by $i$ will split into more than three sub-groups? 

However, only the state-of-the-art approach, mFLICA~\cite{mFLICASDM18}, is capable of inferring dynamics of leadership -- i.e. emergence, convergence, or a change of leaders -- during coordinated movement.  mFLICA detects clusters (factions) based on the concept of following relations. In mFLCIA, the time series from the same faction must follow the same leader.

\begin{table}[]
\caption{Comparison of frameworks that can detect clusters in time series. Static clusters are clusters that are defined over data points in each time step, while temporal clusters are defined over segments of time series.  Factions are temporal clusters that all members follow the same leader. Tracking clusters dynamics implies that a framework can track evolution of clusters, such as merging or splitting of clusters over time. The following relation property implies that a framework can give a relation of who follows whom for all pairs of members within a cluster.}
\label{tab:methodComp}
\begin{small}
\begin{tabular}{|c|c|c|c|}
\hline
Properties\textbackslash Frameworks & MONIC                             & TRACLUS,TCMM            & mFLICA                  \\ \hline
Cluster types                       & Static clusters                   & Temporal clusters        & Factions                \\ \hline
Members of clusters                 & Data points  &  Time series & Time series \\ \hline
Tracking clusters dynamics        & Yes                               & No                      & Yes                     \\ \hline
Following relations                 & No                                & No                      & Yes                     \\ \hline
\end{tabular}
\end{small}
\end{table}

There are many works focusing on inferring dynamics of groups or  clusters~\cite{10.1007/978-3-642-12098-5_3,lee2007trajectory,Spiliopoulou:2006:MMM:1150402.1150491}. The work by \cite{Spiliopoulou:2006:MMM:1150402.1150491} proposed a framework named ``MONIC" to track various types of clusters transition in time series, such as expanding, splitting, merging, etc. However, MONIC infers clusters based on time points without considering following relations among time series to detect clusters.  The work by \cite{10.1007/978-3-642-12098-5_3,lee2007trajectory} proposed frameworks (TRACLUS and TCMM) to detect temporal clusters from segments of time series. Nevertheless, the temporal clusters are measured based on trajectory similarity without the following relation property. Hence, MONIC, TRACLUS, and TCMM frameworks cannot be used to detect factions of time series, which implies that they cannot detect leadership and followership dynamics. Table~\ref{tab:methodComp} summarizes the comparison of these frameworks.

In this paper, we focus on mining frequent patterns of leadership dynamics that requires our framework to identify both groups and leaders of those groups that change over time. Moreover, since the groups following a leader during coordination have a special structure of following relations among members, the standard clustering methods cannot be used in this case.   \\

\begin{framed}  
\noindent { \slip :} {  { Given time series of individual activities, the goal is to mine and model frequent patterns of leadership dynamics, including emergence of multiple leaders, convergence of multiple leaders to a single one, or change of a leader.}}
\end{framed}

\subsection{Previous contributions: leadership dynamics}
To address these computational questions, in the previous paper in~\cite{ASONAM18}, we formalize the problem of \slip,  as well as propose a framework, which is the extension of mFLICA~\cite{mFLICASDM18}, as a solution to this problem. We adapt the traditional framework of  frequent pattern mining~\cite{Agrawal:1993:MAR:170035.170072,Han2007,aggarwal2014frequent} and the Hidden Markov Model (HMM) approach~\cite{18626} to model the dynamics of frequent patterns of leadership. Our framework is capable of:\\

\squishlist
\item {\bf Mining and modeling frequent patterns of leadership dynamics :} inferring the transition diagram of frequent dynamics of complex leadership events, such as ``a single group lead by $k$ splits into two groups lead by $i$ and $j$". In addition, we infer the probabilities of the transitions between such two events in the diagram.
\item {\bf Evaluating the significance of leadership-event order:} we propose a null model of the dynamics of leadership events and perform hypothesis testing to compare  frequent-pattern model of leadership dynamics inferred from the given input to our proposed null model.
\item {\bf Mining sequence patterns of leadership dynamics:} finding support values for the leadership-dynamics sequences from time series of movement data.
\item {\bf Evaluating the significance of frequencies of leadership event sequences:} we propose a null model of the sequences of leadership events and perform hypothesis testing to compare  the support distribution of leadership event sequences inferred from the given input to our proposed null model.

\squishend

We use several simulated datasets from the work in ~\cite{mFLICASDM18} that cover various types of leadership dynamics for validation, as well as a dataset of trajectories of baboons~\cite{crofoot2015data,Strandburg-Peshkin1358} to demonstrate the application of our framework. To the best of our knowledge, this is the first work to deal with the topic of complex leadership dynamics and there is no comparable method, therefore, we indirectly compare our framework to an enhanced FLOCK method~\cite{andersson2008reporting}, used as a baseline for leadership inference only. 


\subsection{New contributions: followership dynamics}

While our previous work in~\cite{ASONAM18} addressed several aspects of leadership dynamics, there are still gaps remaining in our understanding of followership dynamics. For instance, assuming individuals $i$ and $j$ are in the same sub-group, how likely is it that they will be in  different groups in the future? How many clusters of individuals are there such that members in each cluster stay together with a support at least 0.7?  How likely is it  for an individual $i$ that its sub-group will  be lead by an individual $j$ from the same group? 

\begin{framed}  
\noindent { \sfip :} {  { Given the time series of individual activities, the goal is to mine and model frequent patterns of followership dynamics, including the change of sub-groups of followers (unity), or the choice of whom to follow (loyalty).}}
\end{framed}

To address these questions, in this paper, we extend our framework to include several aspects of followership dynamics. In addition to the previous work in~\cite{ASONAM18}, our framework is capable of:\\

\squishlist
\item {\bf Mining and modeling frequent patterns of faction membership:} estimating the frequency of a pair of individuals being in the same faction, as well as discovering faction clusters of individuals.
\item {\bf Mining and modeling frequent patterns of leader-follower relationship:} estimating the frequency of each individual being a follower of a group lead by a specific individual, as well as discovering dynamics of the changes of a leader of each faction cluster.
\squishend

We also use simulated datasets from the work in ~\cite{mFLICASDM18} that contain complex leader-follower dynamics to evaluate our framework.  Our approach is flexible to be generalized beyond the time series of movement data to arbitrary time series where subsets intentionally or spontaneously coordinate.

%% file: 03-probstate.tex
\newcommand{\argmax}{\mathop{\mathrm{argmax}}\limits} 
\newcommand{\argmin}{\mathop{\mathrm{argmin}}\limits} 



\section{Problem statement}

In this paper, we use leadership definitions from the work in ~\cite{mFLICASDM18}. Given a $D$-dimensional time series $Q$, we use $Q(t)$ to refer to an element of the time series $Q$ at time $t$ and, for a given $\Delta \in \mathbb{Z}$, $Q_\Delta$ as a time-shifted version of $Q$ where, $Q(t)=Q_\Delta(t+\Delta)$. 

\begin{definition}[$\sigma$-Following relation (\cite{mFLICASDM18})]
\label{followRDef}
Let $\mathcal{U}$ be a set of time series, $\mathrm{sim}: \mathcal{U}\times \mathcal{U} \to [0,1]$ be a time series similarity function, and $\sigma \in[0,1]$ be a similarity threshold. For any $P, Q \in \mathcal{U}$, we say that $Q$  {\em follows} $P$ if $Q$ and $P$ are sufficiently similar within some time shift $\Delta$:
$$\max_\Delta(\mathrm{sim}(P,Q_\Delta) ) \geq \sigma \mbox { and }$$ 
$$ \min(\argmax_\Delta\mathrm{sim}(P,Q_\Delta) \geq 0)\neq \emptyset.$$ 
\end{definition}

Typically, to measure the following relation in Def.~\ref{followRDef}, the work by ~\cite{mFLICASDM18} used the Dynamic Time Warping (DTW) developed by~\cite{sakoe1978dynamic}. DTW is used to measure a distance between two time series. It can measure a distance of multi-dimensional time series (\cite{keogh2005exact}). Since DTW uses Euclidean distance as a kernel to measure a distance between a pair of elements of two time series, the weighted Euclidean distance can be deployed in the case that we want to give some dimensions higher contribution to a distance measure. Additionally, DTW performs better than several methods (\cite{kjargaard2013time}) and robust to the noise (~\cite{doi:10.1137/1.9781611974010.33}) for the task of following relation inference (\cite{Amornbunchornvej:2018:CED:3234931.3201406}). The following relation measure using DTW is bounded in $[0,1]$ interval (Eq.~\ref{eq:traCorr}). In the case that $\mathrm{sim}$ is not bounded, then we need a threshold $\tau$ to normalize the similarity measure. If $\mathrm{sim} \geq \tau$, then the similarity value is one, otherwise zero.

\begin{definition}[Following network (\cite{mFLICASDM18})]
Let $\mathcal{U}$ be a set of time series. A digraph $G=(V,E)$ is a {\em following network} of $\mathcal{U}$ where each node in $V$ corresponds  to a time series in $\mathcal{U}$ and $(Q,P)\in E$ if $Q$ follows $P$.
\end{definition}

\begin{definition}[Initiator of faction (\cite{mFLICASDM18})]
\label{def:initofFaction}
Let $G=(V,E)$ be a following network. $L \in V$ is an {\em initiator of faction} $F_L$ if the out-degree of $L$ is zero and the in-degree of $L$ is greater than zero. The member nodes of $F_L$ are any nodes that have a directed path in $G$ to $L$.
\end{definition}

\subsection{Leadership dynamics}

We create a dynamic following network $\mathcal{G}=\langle{G_t}\rangle$ by considering each temporal sub-interval $t$ of $\mathcal{U}$ of length  $\omega$ (time window parameter) and creating a following network $G_t$ of that interval. We then define the notion of the time series of leaders of a dynamic following network below.

\begin{definition}[Time series of leaders]
Let $\mathcal{U}$ be a set of time series. $\mathcal{L}$ is a {\em time series of leaders} where $\mathcal{L}(t)$ is a set of faction initiators at time $t$ in $G_t$.
\end{definition}

We can use mFLICA framework~\cite{mFLICASDM18} to extract a time series of leaders from time series of movement. Next, we define the support of a leader set $S$. Let $T$ be the length of the time series of leaders and $\mathbbm{1}_x$ be an indicator function, which is $1$ if the statement $x$ is true, and $0$ otherwise. 

\begin{equation}
	\mathrm{supp}_{\mathcal{L}}(S)=\frac{ \sum_{t=1}^T \mathbbm{1} _{ S = \mathcal{L}(t)} }{T}.
	\label{eq:supp}
\end{equation}

$\mathrm{supp}_\mathcal{L}(S)$ indicates the support of having a particular set of initiators $S$ lead multiple groups concurrently. For example, if $\mathrm{supp}_\mathcal{L}(\{L_1,L_2\})=0.5$ it means that half the time the leaders are exactly $\{L_1, L_2\}$, leading their factions concurrently. 

\begin{definition}[Frequent-leader set]
Let $\mathcal{L}$ be a time series of leaders, $S$ be a set of faction initiators, and $\phi\in[0,1]$ be a support threshold. $S$ is a frequent-leader set of ${\mathcal{L}}$ if $\mathrm{supp}_{\mathcal{L}}(S)\geq\phi$.
\end{definition}
\begin{definition}[Transition probability of leader sets]
Let $\mathcal{L}$ be a time series of leaders, and $S_i,S_j$ be sets of faction initiators. A transition probability of leader sets $\lambda_{S_i,S_j}$ is a probability that $\mathcal{L}(t-1)=S_i$ and $\mathcal{L}(t)=S_j$.
\end{definition}
Now, we are ready to formally state the the problem of \slip.

\begin{problem1}
    \SetKwInOut{Input}{Input}
    \SetKwInOut{Output}{Output}
    \Input{A set  $\mathcal{U} = \{U_1,\dots, U_n\}$ of $m$-dimensional time series and a support threshold $\phi$.}
    \Output{A set of frequent-leader sets $\mathcal{S}_\mathcal{L}$ and a transition probability set  $\mathcal{P}=\{\lambda_{S_i,S_j}\}$ where $S_i,S_j\in  \mathcal{S}_\mathcal{L}$. }
    \caption{{\slip}}
    \label{prob:SLIP}
\end{problem1}

In this paper, we choose to represent a set of frequent-leader sets as a diagram of leadership dynamics below.
\begin{definition}[A diagram of leadership dynamics]
\label{def:SMleader}
Let $\mathcal{L}$ be a time series of leaders, $\phi\in[0,1]$ be a support threshold, and $S_\mathcal{L}$ be set of frequent-leader sets. A digraph $\mathcal{T}=(V_\mathcal{T},E_\mathcal{T})$ is a diagram of leadership dynamics such that the nodes $V_\mathcal{T}$ represent frequent-leader sets $S_\mathcal{L}$ and $(v_i,v_j)\in E_\mathcal{T}$ if $\lambda_{S_i,S_j}>0$.
\end{definition}

\subsection{Followership dynamics}

Given a dynamic following network $\mathcal{G}=\langle{G_t}\rangle$ of a set of time series $\mathcal{U}$, we define the notion of the {\em time series of factions} of the dynamic following network below.

\begin{definition}[Time series of factions]
Let $\mathcal{U}$ be a set of time series and $\mathcal{G}=\langle{G_t}\rangle$ be a dynamic following network of $\mathcal{U}$, where $G_t=(V_t,E_t)$ is a following network at time $t$. We say that $\mathcal{F}$ is a {\em time series of factions}, where $\mathcal{F}(t)$ is a set of factions $\{F_L\}$ at time $t$ in $G_t=(V_t,E_t)$ and $F_L\subseteq V_t$ is a set of faction members of initiator $L$  (Def.~\ref{def:initofFaction}).
\end{definition}

The time series of factions contains the information of who belongs to which specific faction over time. Having defined the time series of factions, we can formalize the concept of a pair of individuals who frequently stay together in the same faction.

\begin{definition}[frequent co-faction pair]
\label{def:cofactpair}
Let $\mathcal{F}$ be a time series of factions, $\phi_{CO}$ be a threshold, and $V=\{1,\dots,n\}$ be a set of individual indices. We say that individuals $i$ and $j$ in $V$ are a {\em frequent co-faction pair} if the frequency of  $i$ and $j$ being in the same faction of $\mathcal{F}$ is greater than $\phi_{CO}$.
\end{definition}

A frequent co-faction pair might indicate friendship or other strong affiliation between individuals. At the group level, we define the notion of a frequent co-faction cluster below.  

\begin{definition}[frequent co-faction cluster]
\label{def:cluster}
Let $\mathcal{F}$ be a time series of factions, $\phi_{CO}$ be a threshold, and $V=\{1,\dots,n\}$ be a set of individual indices. We say that a set $C\subseteq V$ is a {\em frequent co-faction cluster} if every member pair $i,j$ in $C$ is a frequent co-faction pair with respect to $\mathcal{F}$ and $\phi_{CO}$ and there is no other frequent co-faction cluster $C'\subseteq V$ where $C\subset C'$. In other words, $C$ is a maximal set of frequent co-faction pairs. 
\end{definition}

A frequent co-faction cluster represents a concept of cohesion. If an entire group has a strong level of cohesion, then there are a few clusters. In contrast, if a group has a weak level of cohesion, then there are multiple clusters. The members within the same cluster might either be loyal to a specific leader, share the same interests, or be a group of friends or other strong affiliates. 

 In general, given a time series of factions as an input, the problem of finding global faction clusters of individuals is NP-hard, and an approximated solution was given by~\cite{Tantipathananandh:2007:FCI:1281192.1281269}. However, in our setting, a faction or cluster is a group of followers lead by a particular frequent leader. Hence, the problem of finding global faction clusters reduces to the problem of finding followers of each frequent leader, which can be done in polynomial time in one scan over the time series.   

To illustrate the concept of loyalty of a follower toward a specific leader, we formalize the notion of a {\em frequent leader-follower pair} below.

\begin{definition}[frequent leader-follower pair]
\label{def:leadFollpair}
Let $\mathcal{F}$ be a time series of factions, $\phi_{LF}$ be a threshold, and $V=\{1,\dots,n\}$ be a set of individual indices. We define the order pair ($i,j$) in $V$ is the {\em frequent leader-follower pair} if the frequency of having $i$ within a faction that has $j$ as the initiator with respect to $\mathcal{F}$ is greater than $\phi_{LF}$.
\end{definition}

A frequent leader-follower pair $i,j$ implies that $i$ is a member of a faction lead by $j$ most of the time. This implies that $i$ might be a loyal follower of $j$ (or have a strong affiliation with a loyal follower of $j$).

Now, we are ready to formally state the the problem of \sfip.
\begin{problem1}
    \SetKwInOut{Input}{Input}
    \SetKwInOut{Output}{Output}
    \Input{A set  $\mathcal{U} = \{U_1,\dots, U_n\}$ of $m$-dimensional time series, threshold $\phi_{CO}$, and threshold $\phi_{LF}$.}
    \Output{A set of co-faction pairs $\mathcal{S}_{CO}$, a set of frequent co-faction clusters $\mathcal{S}_{CL}$, and a set of  frequent leader-follower pair $\mathcal{S}_{LF}$. }
    \caption{{\sfip}}
    \label{prob:SFIP}
\end{problem1}

In this paper, we choose to represent a set of frequent co-faction pairs as a co-faction network, as well as choose to represent a set of frequent leader-follower pairs as a lead-follow network below.

\begin{definition}[A co-faction network]
\label{def:cofacNet}
Let $\mathcal{F}$ be a time series of factions, $\phi_{CO}$ be a threshold, and $V=\{1,\dots,n\}$ be a set of individual indices. An undirected graph $\mathcal{G}_{CO}=(V,E_{CO})$ is a  co-faction network such that there is an edge $(v_i,v_j)\in E_{CO}$ if a pair $(v_i,v_j)$ is a frequent co-faction pair w.r.t.  $\mathcal{F}$ and $\phi_{CO}$. The weight of the edge $(v_i,v_j)$ is a frequency of having $i,j$ in the same faction.
\end{definition}

\begin{definition}[A lead-follow network]
\label{def:leadfollNet}
Let $\mathcal{F}$ be a time series of factions, $\phi_{LF}$ be a threshold, and $V=\{1,\dots,n\}$ be a set of individual indices. A bipartite graph $\mathcal{G}_{LF}=(V_F,V_L,E_{LF})$ is a lead-follow network where $V_F$ represents a set of follower nodes, and $V_L$ represents a set of initiator nodes. For any $v_i\in V_F$ and $v_j \in V_L$, there is a directed edge $(v_i,v_j)\in E_{LF}$ if an order pair $(v_i,v_j)$ is a frequent leader-follower pair w.r.t.  $\mathcal{F}$ and $\phi_{LF}$. 
\end{definition}

%% file: 04-method.tex
\label{sec:methods}
\begin{figure*}[!ht]
\centering
\includegraphics[width=\textwidth]{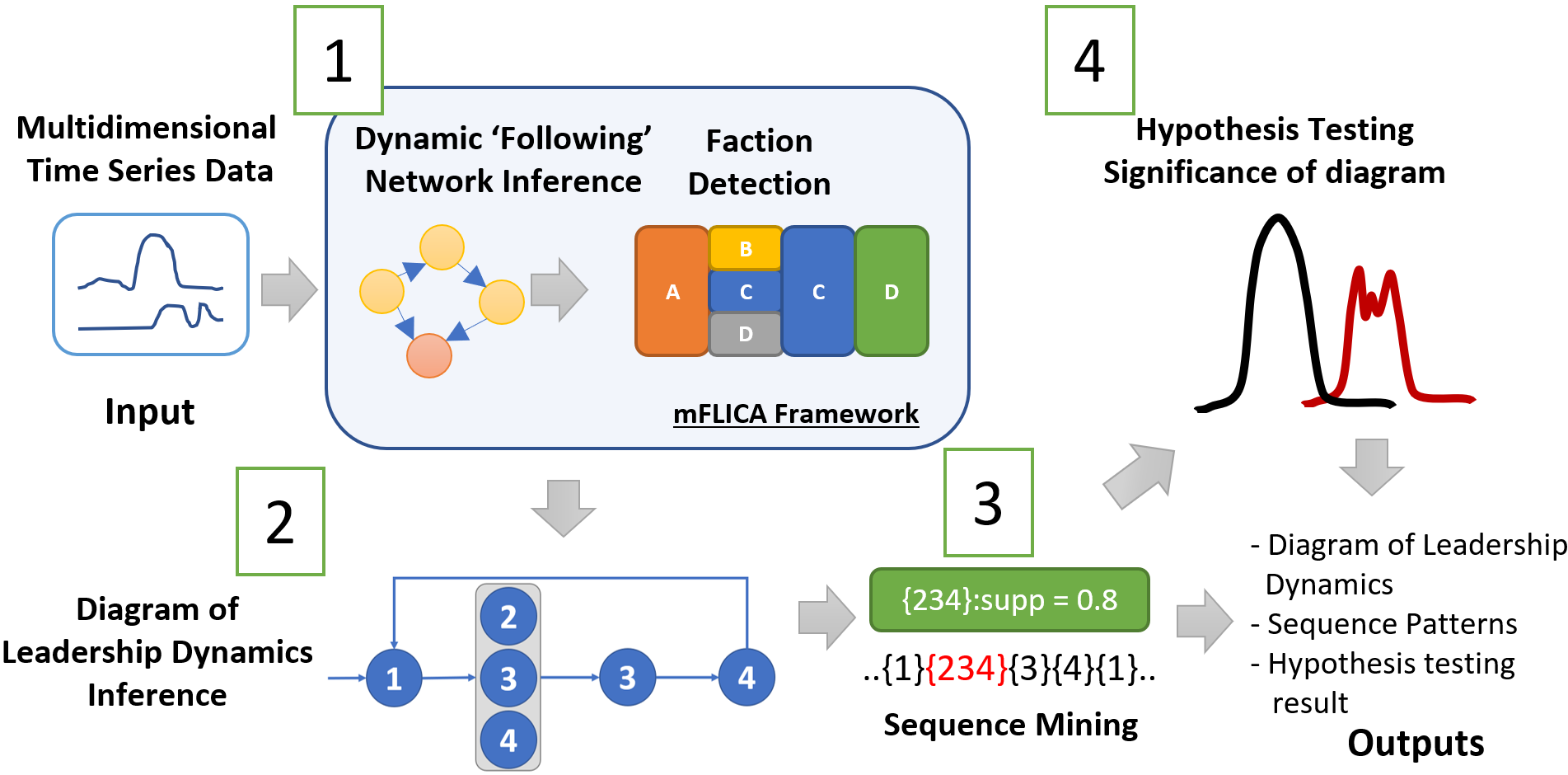}
\caption{A high-level overview of the proposed framework for inferring leadership dynamics. }
\label{fig:SMOverview}
\end{figure*}

\section{Methods}
\subsection{Leadership dynamics}
To solve Problem~\ref{prob:SLIP}, we propose the framework consisting of four parts (Fig.~\ref{fig:SMOverview}).  Given a set of time series of movement $\mathcal{U}=\{U_1,\dots,U_n\}$, where $U_i \in \mathcal{U}$ is a two-dimensional time series of length $T$, first, we infer a dynamic following network and time series of leaders $\mathcal{L}$ using mFLICA framework~\cite{mFLICASDM18} (Section~\ref{sec:mFLICA}). Second, we infer a diagram of leadership dynamics $\mathcal{T}$ from $\mathcal{L}$ in Section~\ref{sec:SMinfermethod}. Third, we detect the sequence patterns on $\mathcal{L}$ in Section~\ref{sec:Miningmethod}. Finally, we deploy hypothesis tests to evaluate significance of leadership dynamics compared to our proposed null models in Section~\ref{sec:Htestmethod}.  
\subsection{mFLICA}
\label{sec:mFLICA}
Given a pair of time series $U$ and $Q$, mFLICA uses Dynamic Time Warping (DTW)~\cite{sakoe1978dynamic} to infer a following relation. Suppose $P_{U,Q}$ is an optimal warping  path from DTW dynamic programming matrix, where $(i,j)\in P_{U,Q}$ implies $U(i)$ matched with $Q(j)$ in the matrix. Intuitively, if $U$ is followed by $Q$ with the time delay $\Delta_{i,j}$, then $j-i=\Delta_{i,j}$. Hence, we can compute the {\em following} relation by the equation below.

\begin{equation}
	\mathrm{f}(P_{U,Q})=\frac{\sum_{(i,j) \in P_{U,Q}}\mathrm{sign}(j-i)}{|P_{U,Q}|}.
	\label{eq:traCorr}
\end{equation}

Suppose we have a similarity threshold $\sigma$, there we say there is a following relation if $|\mathrm{f}(P_{U,Q})| \geq\sigma$, where $Q$ follows $U$ if $\mathrm{f}(P_{U,Q})\geq \sigma$ and $U$ follows $Q$ if $\mathrm{f}(P_{U,Q})\leq -\sigma$. We set $\sigma =0.5$ as a default.

Next, given  a time window  $\omega$  and a sliding window parameter $\delta=0.1\omega$, we have the $i^\mathrm{th}$ time window interval $w(i) = [(i-1)\times\delta,(i-1)\times\delta+\omega]$. mFLICA creates a following network for each set of time series within interval $w(i)$ of $\mathcal{U}$. An edge of a following network is inferred by Eq.~\ref{eq:traCorr} with the weight $|\mathrm{f}(P_{U,Q})|$. Hence, after every interval $w(i)$ has its following network, we have a dynamic following network $\mathcal{G}=\langle{G_t}\rangle$ of $\mathcal{U}$.

Lastly,  for each time step $t$, mFLICA uses Breadth First Search (BFS) to infer factions and initiators within a following network $G_t$. The  faction initiators are nodes with out-degree zero and  in-degree non-zero. By applying BFS to dynamic following network $\mathcal{G}$, we have the time series of leaders $\mathcal{L}=(\mathcal{L}(1),\dots,\mathcal{L}(T))$ as well as the time series of factions $\mathcal{F}=(\mathcal{F}(1),\dots,\mathcal{F}(T))$ as the outputs of this step.

\subsection{Inferring transition diagram of leadership dynamics}
\label{sec:SMinfermethod}
We use Hidden Markov Model (HMM)~\cite{18626} to model a diagram of leadership dynamics $\mathcal{T}=(V_\mathcal{T},E_\mathcal{T})$ in Def.~\ref{def:SMleader} and use Baum--Welch algorithm~\cite{jelinek1975design} to infer the maximum likelihood estimates of parameters of HMM from the time series of leader $\mathcal{L}$.  In this setting, we have a set of frequent-leader sets  $\mathcal{S}_\mathcal{L}$ as a set of states  in HMM with the support threshold $\phi=0.01$. In HMM, the stochastic transition matrix $A$, which has its size $|\mathcal{S}_\mathcal{L}|\times |\mathcal{S}_\mathcal{L}|$, describes estimated probabilities that a group changes its current set of leaders to another set of leaders (e.g. group merging or splitting.)  However, since we are interested only in the events of state changes, we ignore the self-transition probability and normalize $A$ to be $A^*$ (Eq.~\ref{eq:Anormalize}), which is the adjacency matrix of $\mathcal{T}$.

Given a time series of leaders $\mathcal{L}$, we can easily infer  a set of leader sets $\mathcal{S}_\mathcal{L}$. Then, let $\mathcal{S}_\mathrm{HMM}$ be  a set of states  in HMM where $\mathcal{S}_\mathrm{HMM}$ and $\mathcal{S}_\mathcal{L}$ are in one-to-one correspondence. We represent each state in $S_\mathrm{HMM}$ as a number in $[1,|\mathcal{S}_\mathrm{HMM}|]$, then we create $\mathcal{L}_\mathrm{HMM}$ by replacing each element in $\mathcal{L}$ with the number of corresponding state in $\mathcal{S}_\mathrm{HMM}$. For example, in Fig.~\ref{fig:Dytype} (Dynamic Type 1), we have  a set of leader sets $\mathcal{S}_\mathcal{L}$ and $\mathcal{S}_\mathrm{HMM}$, where $\mathrm{\{ID1\},\{ID2,ID3,ID4\},\{ID3\}}$, and $\mathrm{\{ID4\}}$, in $\mathcal{S}_\mathcal{L}$ have corresponding elements in  $\mathcal{S}_\mathrm{HMM}$ as $1,2,3$ and $4$, respectively.

Initially, we set a stochastic transition matrix $A=\{a_{i,j}\}$ ($i,j$ are the states) and the initial state distribution $\pi_i$ uniformly. We have the set of observation values $Y=\{1,\dots,|\mathcal{S}_\mathrm{HMM}|\}$. In this setting, there is no hidden state since an observation value is an identity of a state. However, in HMM, at any state $i$, there is a required probability $b_{i,j}$ of observing value $j$ at the state $i$ (typically represented by a matrix $B=\{b_{i,j}\}$.)  Here, the probability $b_{i,j}=1$ if $i=j$ and zero otherwise.

We use Baum--Welch algorithm~\cite{jelinek1975design} to infer $A=\{a_{i,j}\}$, then we normalize $A$ to create $A^*=\{a^*_{i,j}\}$ by the equation below. 

\begin{equation}
  	a^*_{i,j}=\left\{
  \begin{array}{@{}ll@{}}
	0, & i=j \\
   \frac{a_{i,j}}{\sum_{k=1,k\neq j}^{|S_\mathrm{HMM}|} a_{i,k} }, & \mathrm{Otherwise}. \\
  \end{array}\right.
\label{eq:Anormalize}
\end{equation}

\subsection{Mining sequence patterns of leadership dynamics}\label{sec:Miningmethod}
After having a diagram of leadership dynamics $\mathcal{T}=(V_\mathcal{T},E_\mathcal{T})$, for each pair of nodes $(i,j)\in V_\mathcal{T}$, we find a sequence pattern, which is a path $P_{i,j}=(v(1)=i,\dots,v(k)=j)$, where for all $u \in  V_\mathcal{T}$, $a^*_{v(t-1),v(t)} > a^*_{v(t-1),u}$.

$P_{i,j}$ is an order sequence that the previous state $v(t-1) \in P_{i,j}$ has the highest probability to change to the next consecutive state $v(t) \in P_{i,j}$, given a starting point at $i$ and the final state at $j$.

Given $A^*=\{a^*_{i,j}\}$ as an adjacency matrix of $\mathcal{T}$, we convert $A^*$ to be $A'=\{a'_{i,j}\}$ where $a'_{i,j}=\frac{1}{a^*_{i,j}}$. Then, we use the standard Dijkstra's algorithm to find the shortest path between every two nodes in $A'$. Hence, $P_{i,j}$ is the shortest path between $i$ and $j$ in $A'$\footnote{Note, this can be done since the probability condition is independent of each pair and not cumulative over the path}.  Let $\nu$ be a number of times that the full sequence of $P_{i,j}$ occurs in $\mathcal{L}$ and $N$ be a number of times that leadership state change happens in $\mathcal{L}$ ({\em e.g.}, two sub-groups merged together, changing the leader), we can find the support of  $P_{i,j}$ in  the time series of leader $\mathcal{L}$ by the equation below: 

\begin{equation}
	\mathrm{supp}_{path}(\mathcal{L},P_{i,j})=\frac{\nu\times(|P_{i,j}|-1)}{N}.
	\label{eq:suppath}
\end{equation}

 Specifically, $\nu$ is a number of times that all pairs of nodes $v(t-1),v(t) \in P_{i,j}$ s.t. $v(t-1)$ appear before $v(t)$ in $P_{i,j}$  also appear in $\mathcal{L}$.

\subsection{Hypothesis testing}
\label{sec:Htestmethod}

\begin{table}[]
\centering
\caption{Details of  non-parametric tests used in this paper. A significant level has been set at $\alpha=0.01$ for all experiments.}
\label{table:htestdes}
\begin{tabular}{|c|c|}
\hline
\rowcolor[HTML]{EFEFEF} 
\textbf{Method}                                                          & \textbf{Null hypothesis $H_0$}                  \\ \hline
Kolmogorov-Smirnov test~\cite{doi:10.1080/01621459.1951.10500769} & Two samples are from       \\ \cline{1-1}
Wilcoxon Rank Sum Test~\cite{10.2307/3001968}                     & the same distribution                    \\ \cline{1-1}
Kruskal-Wallis Test~\cite{doi:10.1080/01621459.1952.10483441}     &                      \\ \hline
\end{tabular}
\end{table}

\subsubsection{Evaluating the significance of leadership-event order}
\label{sec:sig1}
 Given a time series of leaders $\mathcal{L}$ and a diagram of leadership dynamics  $\mathcal{T}$ inferred from $\mathcal{L}$, we perform a random permutation of elements in $\mathcal{L}$ to create $\mathcal{L}_{\mathrm{rand}}$, then we infer a diagram of leadership dynamics  $\mathcal{T}_{\mathrm{rand}}$ from $\mathcal{L}_{\mathrm{rand}}$ by the method described by the previous section. 
Afterwards, we test the similarity of the edge-weight distributions of $\mathcal{T}$ and $\mathcal{T}_{\mathrm{rand}}$. We deploy three non-parametric methods, shown in Table~\ref{table:htestdes}, to perform the tests. If all three methods successfully reject the null hypothesis with the significant threshold $\alpha=0.01$, then we conclude that the edge-weight distribution of $\mathcal{T}$ is significantly different from  $\mathcal{T}_{\mathrm{rand}}$'s although the support value of each node in both graphs are the same. 

\subsubsection{Evaluating the significance of frequencies of leadership-event sequences}
\label{sec:sig2}
After finding all the sequences for every pair of nodes in Section~\ref{sec:Miningmethod}, we compute the support $\mathrm{supp}_{path}(\mathcal{L},P_{i,j})$ of each sequence $P_{i,j}$. This gives the sequence-support distribution of $\mathcal{T}$. Next, we rewire $\mathcal{T}$ to be $\mathcal{T}_{\mathrm{rand}}$ by uniformly and randomly changing the end points of each edge in $\mathcal{T}$, then we calculate the sequence-support distribution of $\mathcal{T}_{\mathrm{rand}}$ (Eq.~\ref{eq:suppath}.) Lastly, we test whether $\mathcal{T}$ and $\mathcal{T}_{\mathrm{rand}}$  sequence-support distributions are different the same way as in the previous section.

We repeat both types of significance tests 100 times and report the percentage of times that the tests successfully reject $H_0$ for each dataset. 

\subsection{Followership dynamics}

\begin{figure}[!ht]
\centering
\includegraphics[width=\textwidth]{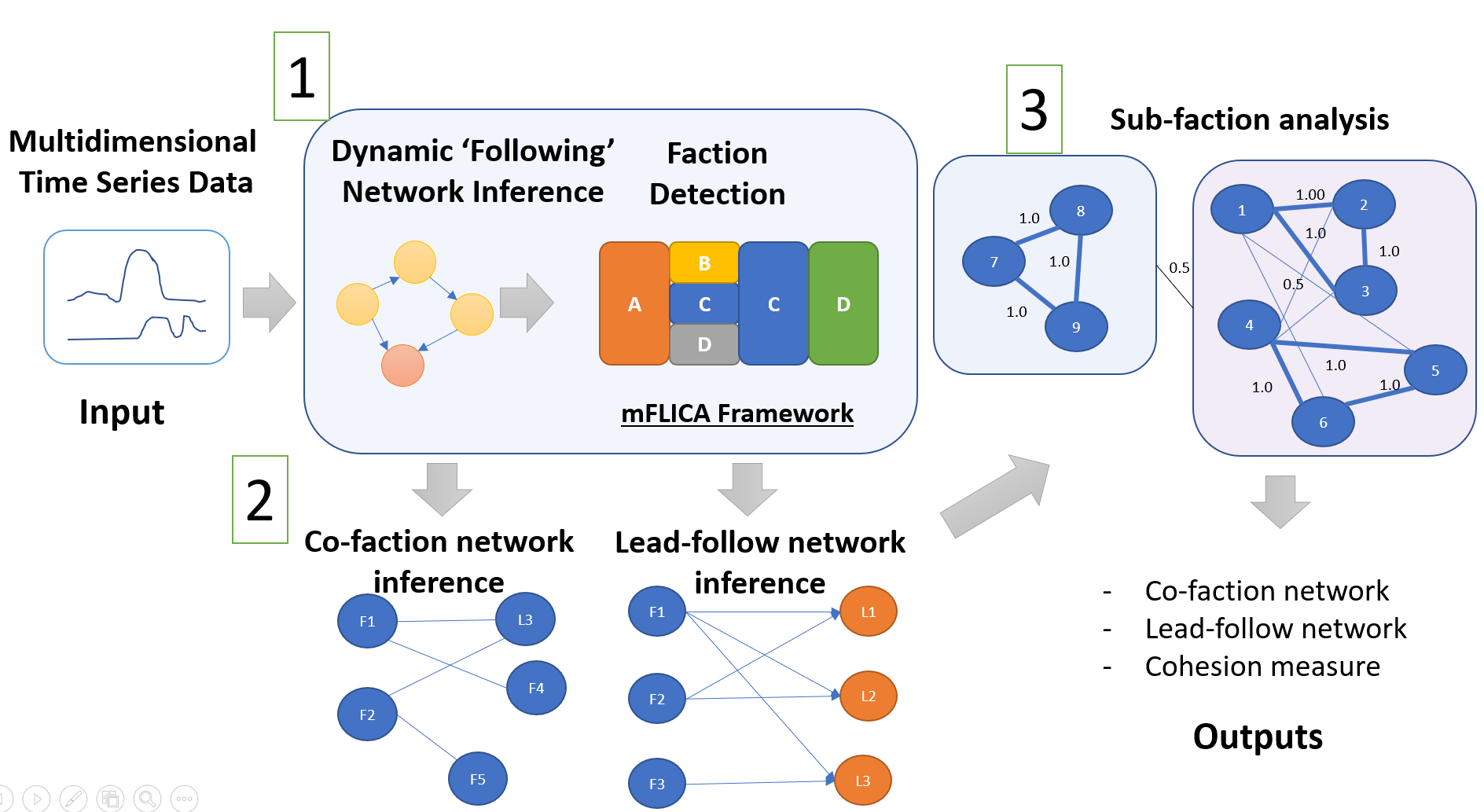}
\caption{A high-level overview of the proposed framework for inferring followership dynamics. }
\label{fig:FOLLOverview}
\end{figure}

To solve Problem~\ref{prob:SFIP}, we propose the framework that consists of three parts (Fig.~\ref{fig:FOLLOverview}).

Let $\mathcal{U}=\{U_1,\dots,U_n\}$ be a set of time series of movement, where $U_i \in \mathcal{U}$ is a time series of length $T$. In the first step, we infer a dynamic following network as well as a time series of factions $\mathcal{F}$ using mFLICA framework~\cite{mFLICASDM18} (Section~\ref{sec:mFLICA}).
  
Second, we infer a co-faction network $\mathcal{G}_{CO}=(V,E_{CO})$ (Section~\ref{sec:cofactnetInfer}) and a lead-follow network $\mathcal{G}_{LF}=(V,E_{LF})$ (Section~\ref{sec:leadfollnetInfer}) from $\mathcal{F}$. Afterwards, we infer a set of frequent co-faction clusters $\{C\}$ (Section~\ref{sec:subfac}).
  
\subsubsection{Co-faction network inference}
\label{sec:cofactnetInfer}
To infer a co-faction network, the first step is to infer a pair of frequent co-faction in Def.~\ref{def:cofactpair}.

Given a time series of factions $\mathcal{F}$ with the length $T$ and an indicator function $\mathbbm{1}_x$ (which is $1$ if the statement $x$ is true, and $0$ otherwise), we define the support of having individuals $i$ and $j$ in the same faction below. 
\begin{equation}
	\mathrm{csupp}_{\mathcal{F}}(i,j)=\frac{ \sum_{t=1}^T \mathbbm{1} _{\exists F\in \mathcal{F}(t), \{i,j\} \subseteq F} }{T}.
	\label{eq:csupp}
\end{equation}

Here $\mathrm{csupp}_{\mathcal{F}}(i,j)$ indicates the support of having a particular pair of individuals $i$ and $j$ being within the same faction in $\mathcal{F}$. 

After we compute the supports $\mathrm{csupp}$ for all pairs of individuals, we have a co-faction network $\mathcal{G}_{CO}=(V,E_{CO})$. Given a threshold $\phi_{CO}$, there is an edge $(v_i,v_j)\in E_{CO}$ if $\mathrm{csupp}_{\mathcal{F}}(i,j)\geq \phi_{CO}$. The edge weight of  $(v_i,v_j)$ is  $\mathrm{csupp}_{\mathcal{F}}(i,j)$. 

\subsubsection{Lead-follow network inference}
\label{sec:leadfollnetInfer}
To infer a lead-follow network, the first step is to infer a frequent leader-follower pair $i,j$ in Def.~\ref{def:leadFollpair}. Given a time series of factions $\mathcal{F}$ with the length $T$ and an indicator function $\mathbbm{1}_x$ (which is $1$ if the statement $x$ is true, and $0$ otherwise), we can define a support of having individual $i$ in the faction lead by an initiator $j$ below.

\begin{equation}
	\mathrm{lfsupp}_{\mathcal{F}}(i,j)=\frac{ \sum_{t=1}^T \mathbbm{1} _{\exists F_{j}\in \mathcal{F}(t), i \in F_j} }{T}.
	\label{eq:LFsupp}
\end{equation}

Where $F_j$ is a set of faction members leading by $j$.  Here $\mathrm{lfsupp}_{\mathcal{F}}(i,j)$ indicates the support of having a particular individual $i$ in the faction leading by an initiator $j$ in $\mathcal{F}$.  

After we compute the supports $\mathrm{lfsupp}$ for all pairs of individuals, then we have a lead-follow network $\mathcal{G}_{LF}=(V_F,V_L,E_{LF})$. Given a threshold $\phi_{LF}$, for any  $i\in V_F$ and $j \in V_L$, there is a directed edge $(v_i,v_j)\in E_{CO}$ if $\mathrm{lfsupp}_{\mathcal{F}}(i,j)\geq \phi_{LF}$. The edge weight of  $(v_i,v_j)$ is  $\mathrm{lfsupp}_{\mathcal{F}}(i,j)$.  Higher $\mathrm{lfsupp}_{\mathcal{F}}(i,j)$ implies that there is a higher frequency that $i$ is a member of $j$'s faction. Hence, we can use $\mathrm{lfsupp}_{\mathcal{F}}(i,j)$ as a proxy of loyalty of $i$ to $j$. Higher $\mathrm{lfsupp}_{\mathcal{F}}(i,j)$ implies that $i$ is more loyal to $j$. 

\subsubsection{Clustering and cohesion measure}
\label{sec:subfac}
We use the standard Hierarchical clustering with shortest distance to link clusters~\cite{doi:10.1093/comjnl/16.1.30} to demonstrate our framework ability. However, any clustering algorithm can be used in our framework to perform the analysis.  The  Hierarchical clustering algorithm is an agglomerative clustering approach that starts with each individual in a cluster by itself. Then, it keeps merging two closest clusters to be a single new cluster. The algorithm keeps merging on a set of clusters until there is only a single cluster. Given $C$ and $C'$ are clusters and $ADJ_{CO}$ is an adjacency matrix of a co-faction network that has its element as $\mathrm{csupp}_{\mathcal{F}}(i,j)$, the distance between two clusters is defined below.    

\begin{equation}
	\mathrm{dist}_\mathrm{SingleLink}(C,C')=\min_{i\in C, j \in C'}(dist(ADJ_{CO}(i,*),ADJ_{CO}(j,*))).
	\label{eq:SingleLinkDist}
\end{equation}

Where $ADJ_{CO}(i,*)$ represents an $i$th vector row of $ADJ_{CO}$ and $dist()$ is a standard euclidean distance. The reason that we compute the distance between the vector of weights of $i$ to all individuals and the vector of weights of $j$ to all individuals in $\mathrm{dist}_\mathrm{SingleLink}(C,C')$ is that because two individuals who share the same set of $\mathrm{csupp}_{\mathcal{F}}(i,j)$ are likely members of the same faction. Hence, they should have a small distance. 

Next, since there are two types of edges in  $ADJ_{CO}$: edges that connect members within the same clusters and edges that connect individuals of different clusters. We can use k-means algorithm where $k=2$ to cluster a list of edge weight of the hierarchical tree into two types: internal edges and external edges. Finally, we link any leaves (individuals) of hierarchical tree that are reachable using internal edges to be a member of the same group to represent a frequent co-faction cluster in Def~\ref{def:cluster}.  

To measure the degree of cohesion of $ADJ_{CO}$, we use the standard Modularity Measure (Q value) proposed by M. E. J. Newman and M. Girvan (2004)~\cite{PhysRevE.69.026113} below.

\begin{equation}
	\mathrm{Q}(ADJ,\mathcal{C})=\sum_{c=1}^{|\mathcal{C}|}( e_{i,i} - a^2_i),
	\label{eq:Qeq}
\end{equation}
 where $e_{i,j}$ is a fraction of edges that have one end connected to a node in a cluster $i$ and another end connected to a member of a cluster $j$, and  $a_i=\sum_j e_{i,j}$.  The value of $\mathrm{Q}$ has a range between $-1$ and $1$. If the value is a large positive, then there are multiple strong clusters; the numbers of edges within groups are greater than the numbers of edges between groups.   When there are multiple subgroups that have higher edge weight within the same cluster while edges that connect nodes from different clusters have lower edge weights, then  $\mathrm{Q}$ is close to one. In contrast, if either there is only one cluster or edge weights of all pairs of nodes are not different from each other, then $\mathrm{Q}$ is  close to zero. In other words, higher $\mathrm{Q}$ implies higher number of subgroups that have relatively high edge-weight between nodes within the same cluster compared to edge-weights of nodes from different clusters. Hence, we can use $\mathrm{Q}$ as a proxy of cohesion of group. Higher $\mathrm{Q}$ implies lower cohesion.
 
\subsection{Time and space complexity}

The time complexity of mFLICA is $\mathcal{O}(n^2 \times \omega \times T)$, where $n$ is a number of time series, $T$ is a length of time series, and $\omega$ is a time window parameter. The time complexity of Baum$-$Welch algorithm to infer a diagram of leadership dynamics is $\mathcal{O}(m^2\times T)$ where $m$ is the number of frequent-leader sets. Typically, $m<n$ since there are fewer frequent-leader sets than individuals. In the followership part, we can scan a time series of factions $\mathcal{F}$ only once to compute everything, which has the time complexity at most  $\mathcal{O}(n \times n \times T)$. Hence, our framework's overall time complexity is $\mathcal{O}(n^2 \times \omega \times T)$.  For the space complexity, the most expensive part of our framework is the space for the dynamic following network, which is $\mathcal{O}(n^2\times T)$.

\subsection{Parameters sensitivity}
For the time window parameter $\omega$, the work by~\cite{Amornbunchornvej:2018:CED:3234931.3201406} reported that the following relation is robust to the noise. However, if we set  $\omega$ below the maximum time delay between time series, then the result can be severely affected. Hence, a user should try to guess the maximum time delay on his/her dataset before setting  $\omega$.  Since the core engine of mFLICA is the following relation measure, it is important to set $\omega$ properly. The other parameter such as significant level $\alpha$ should be fine- tuned w.r.t. the task. 

%% file: 05-experiment.tex
\section{Evaluation Datasets}

\begin{figure}[!ht]
\begin{center}
\includegraphics[width=0.7\columnwidth]{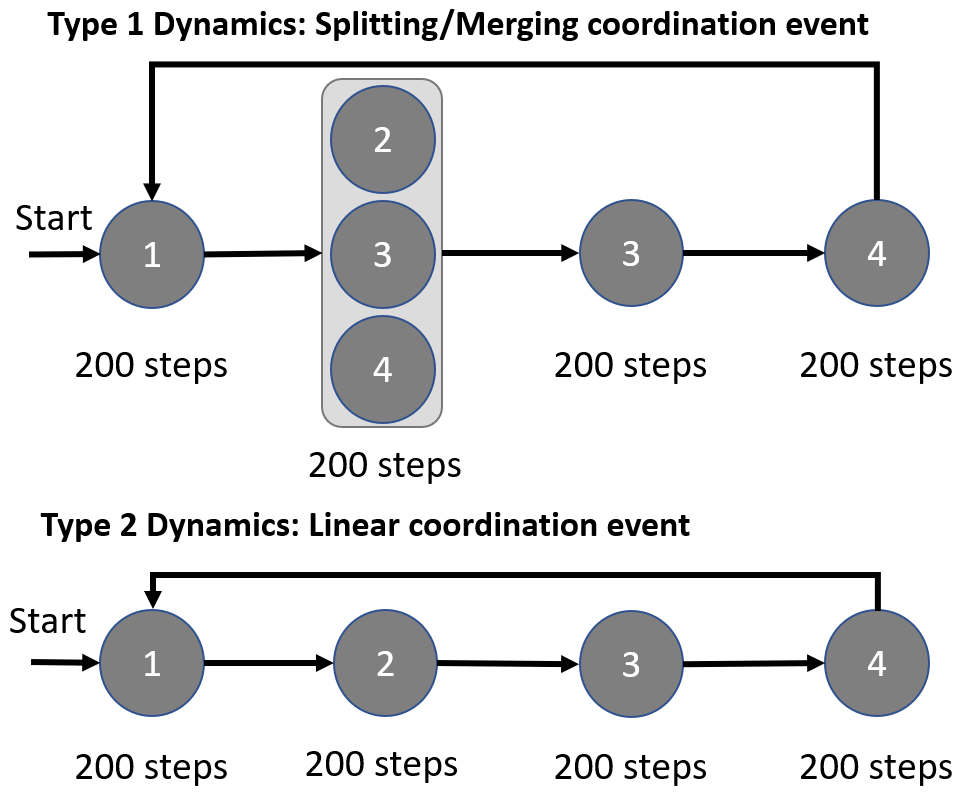}
\end{center}
\caption{Splitting/Merging  (above)  and Linear (below) coordination event. Each node represents the ID of leader of each sub-group at the particular time and each edge represents the change of group's leaders.} 
\label{fig:Dytype}
\end{figure}

We evaluate our method on synthetic datasets generated using a variety of leadership models with a variety of patterns of leadership dynamics.

\subsection{Leadership models}
\label{sec:LeadershipModels}
There are three leadership models that we consider in this paper below.
\subsubsection{Dictatorship Model (DM)~\cite{mFLICASDM18}} Initially, all individuals stay in the initial area. Then, a single initiator moves toward a target path before others. Afterwards, all other individuals follow the initiator with some time delay.

\subsubsection{Hierarchical Model (HM)~\cite{mFLICASDM18}} Each individual has been assigned the unique ranking value at the beginning. Lower rank individuals always follow higher rank individuals. An initiator who has the highest-rank individual (initiator) starts moving first, then the second high-rank individual follows the first-rank individual with sometime delay and so on (the $k+1$th rank individual follows the $k$th- rank individual).  

\subsubsection{Independent Cascade Model (IC)~\cite{kempe2003maximizing}}
Initially, all individuals are deactivated. At the beginning, each individual has a chance to be active with the probability $\rho$. After activation, the active individuals move following the initiator except the initiator itself that follows in the target direction. In every time step, active individuals attempt to activate their $k$-nearest inactive neighbors with the probability of success $\rho$. Active individuals cannot attempt to activate the same individual twice.  In this paper, we determine the parameter space on a combination of :$k \in \{3,5,10\}$ and $\rho \in \{0.25,0.50,0.75\}$.
\subsection{Synthetic trajectory simulation.}
We use simulated datasets to evaluate the performance of our framework. Each dataset consists of 30 individuals. The trajectory of each individual is two-dimensional time series of length 4000 time steps. Each dataset has been generated from one of the three leadership models described above. There are five coordination events in each dataset. One coordination event lasts for 800 time steps. There are two types of coordination events as follows.

\subsubsection{Type 1 Dynamics: Splitting/Merging coordination event~\cite{mFLICASDM18}} 
In this type of coordination event (Fig.~\ref{fig:Dytype} above), ID1 leads the entire group for 200 time steps. Then, the group splits into three equal size sub-groups lead by ID2, ID3, and ID4, for the duration of 200 time steps. Afterwards, all sub-groups are merged into a single group again lead by ID3 for another 200 time steps.  Finally, ID4 leads the entire group for the last 200 time steps. 

\subsubsection{Type 2 Dynamics: Linear coordination event~\cite{mFLICASDM18}} 
In this type of coordination event (Fig.~\ref{fig:Dytype} below), ID1 leads first, then ID2 leads, ID3 leads after ID2, and ID4 leads after ID3. Each leader leads the group for 200 time steps.

After a coordination event ends, then, the group stops moving and the next coordination event repeats the pattern. In this paper, we generated 100 datasets for each leadership model and coordination type (e.g. DM with Type 1 dynamics has 100 datasets). One exception: IC has nine cases of different parameters settings and we have a 100 datasets for each parameter setting and dynamics type. In total, we have 400 datasets for DM and HM but 1800 datasets for IC. 

\subsection{Baboon Dataset}
We also deploy our framework on a dataset of GPS trajectories of wild olive baboons (\emph{Papio anubis}) living at Mpala Research Centre, Kenya \cite{crofoot2015data,Strandburg-Peshkin1358}. The dataset consists of latitude-longitude location time series of 16 baboons recorded for every second for a nine day period (419,095 time steps). We employ this dataset to demonstrate the potential of our framework to uncover relationships within data to generate scientific hypotheses.

\section{Evaluation criteria}
\label{sec:Eval}

\subsection{Leadership dynamics}
In simulated datasets, we compare the inferred adjacency matrix $A=\{a_{i,j}\}$ of a digraph of leadership dynamics $\mathcal{T}=(V_\mathcal{T},E_\mathcal{T})$ against the ground truth matrix $A^*=\{a^*_{i,j}\}$. For the Splitting/Merging coordination event, the ground-truth set of frequent-leader sets is $$\mathcal{S}^*_\mathcal{L}=\mathrm{\{\{ID1\},\{ID2,ID3,ID4\},\{ID3\},\{ID4\}\}}.$$ All elements in $A^*$ are zero except $$a^*_{\mathrm{\{ID1\},\{ID2,ID3,ID4\}}}=a^*_{\mathrm{\{ID2,ID3,ID4\},\{ID3\}}}= a^*_{\mathrm{\{ID3\},\{ID4\}}}= a^*_{\mathrm{\{ID4\},\{ID1\}}}=1.$$ For the Linear coordination event, $$\mathcal{S}^*_\mathcal{L}=\mathrm{\{\{ID1\},\{ID2\},\{ID3\},\{ID4\}\}}$$ and all elements in $A^*$ are zero except $$a^*_{\mathrm{\{ID1\},\{ID2\}}}=a^*_{\mathrm{\{ID2\},\{ID3\}}}= a^*_{\mathrm{\{ID3\},\{ID4\}}}= a^*_{\mathrm{\{ID4\},\{ID1\}}}=1.$$

 Let $\mathcal{S}_\mathcal{L}$ and $\mathcal{S}^*_\mathcal{L}$ be the predicted and the ground truth sets of frequent-leader sets, respectively. The loss function of $A$ and $A^*$ is below:
\begin{equation}
\begin{split}
  	&loss(A,A^*) = \\
  	&\frac{\sum_{i,j \in \mathcal{S}^*_\mathcal{L}\cap \mathcal{S}_\mathcal{L} } |a_{i,j} - a^*_{i,j}|  + \mathrm{FP}(A,A^*)+ \mathrm{FN}(A,A^*)}{n_{A^*}}
\label{eq:lossFunc}
\end{split}
\end{equation}
\begin{equation}
  	 \mathrm{FP}(A,A^*)=\sum_{i,j \in \mathcal{S}_\mathcal{L}\setminus \mathcal{S}^*_\mathcal{L}} |a_{i,j}|
\end{equation}
\begin{equation}
  	 \mathrm{FN}(A,A^*)= \sum_{i, j \in \mathcal{S}^*_\mathcal{L}\setminus \mathcal{S}_\mathcal{L} } |a^*_{i,j}|
\end{equation}
  Where $n_{A^*}$ is the number of elements within $A^*$. The first term in Eq.~\ref{eq:lossFunc} represents the $L1$-norm difference between each element in $A$ and $A^*$ (probabilities) when the predicted states are the same as the ground truth. The second term represents the false positive case when the framework predicts the states that do not exist in the ground truth. The last term represents the false negative case when the framework misses prediction of a state that exists in the ground truth.
  
  \subsection{Followership dynamics}
   \subsubsection{Co-faction network}
  In simulated datasets, we compare an inferred adjacency matrix $A=\{a_{i,j}\}$ of a co-faction network against the ground truth matrix $A^*=\{a^*_{i,j}\}$. All members from the same cluster connected with edges that have the weights $$A^*=\{a^*_{i,j}\}=1,$$ while two nodes from different clusters have the weight $$A^*=\{a^*_{i,j}\}=0.75.$$ 
  For the Splitting/Merging coordination datasets, the ground-truth is that there are three clusters: 
  \begin{eqnarray*}
  C_1&=&\{ID1, ID3, ID5, \dots,ID10\},\\
  C_2&=&\{ID4, ID11, \dots,ID19\},\\
  C_3&=&\{ID2, ID20, \dots,ID30\}.
  \end{eqnarray*}
  
  For Linear coordination datasets, all individuals are in the single cluster.  Given $V$ is a set of nodes of $n$ individuals, we use the absolute loss function to evaluate the difference between predicted $A=\{a_{i,j}\}$ and the ground truth $A^*=\{a^*_{i,j}\}$ below:
  
 \begin{equation}
  	 \mathrm{loss}(A,A^*)=\frac{\sum_{i,j \in V} |a_{i,j} - a_{i,j}|}{ {n \choose 2} }.
\end{equation}

 \subsubsection{Lead-follow network}
 We also compare an inferred adjacency matrix $A=\{a_{i,j}\}$ of a lead-follow network against the ground truth matrix $A^*=\{a^*_{i,j}\}$ of $\mathcal{G}^*_{LF}=(V_F^*,V_L^*,E_{LF}^*)$. In both Splitting/Merging and  Linear coordination datasets, $ID1,ID2,ID3$, and $ID4$ are only initiators. Hence, $V_L^*=\{ID1,ID2,ID3,ID4\}$ and $V_F^*=\{ID1,\dots,ID30\}$. 
 
 For Splitting/Merging coordination datasets, given a leader $L=ID1$, for any $j\in V_F^*, a^*_{L,j}=0.25.$
 
 \squishlist
\item  If  $L=ID2$ and $j\in C_3$, then  $a^*_{L,j}=0.25$, while  $a^*_{L,j'}=0$  for $j\notin C_3$.
\item If  $L=ID3$ and $j\in C_1$, then  $a^*_{L,j}=0.50$, while  $a^*_{L,j'}=0.25$  for $j\notin C_1$.
\item If  $L=ID4$ and $j\in C_2$, then  $a^*_{L,j}=0.50$, while  $a^*_{L,j'}=0.25$  for $j\notin C_2$.
\squishend
 
In Linear coordination datasets, for  $L\in V^*_L$ and $j\in V_F^*$, $a^*_{L,j}=0.25.$ 
 
 Let $V_L$ and $V^*_L$ be the predicted and the ground truth sets of initiators of a lead-follow network respectively, we compare the inferred $A$ and the ground-truth $A^*$ using the loss function below:
\begin{equation}
\begin{split}
  	&loss_{\mathrm{LF}}(A,A^*) = \\
  	&\frac{\sum_{i,j \in V^*_L\cap V_L } |a_{i,j} - a^*_{i,j}|  + \mathrm{FP}_{\mathrm{LF}}(A,A^*)+ \mathrm{FN}_{\mathrm{LF}}(A,A^*)}{n_{A^*}}
\label{eq:lossFuncL}
\end{split}
\end{equation}
\begin{equation*}
  	 \mathrm{FP}_{\mathrm{LF}}(A,A^*)=\sum_{i,j \in V_L\setminus V^*_L} |a_{i,j}|
\end{equation*}
\begin{equation*}
  	 \mathrm{FN}_{\mathrm{LF}}(A,A^*)= \sum_{i, j \in V^*_L\setminus V_L } |a^*_{i,j}|
\end{equation*}
  Where $n_{A^*}$ is the number of elements within $A^*$.
  
 \subsubsection{Clustering evaluation}
 
  For Splitting/Merging coordination datasets, the ground truth of first cluster is $C_1$. The second cluster is $C_2$. The third cluster is $C_3$. For linear coordination datasets, all individuals are in the single cluster. We use $F1$ score to measure the difference between inferred and ground-truth clusters. Given $C_i$ is a ground-truth cluster and $\hat{C}_j$ is a predicted cluster that have the most common members with $C_i$. The true positive is a sum of number of common members between all pair of $C_i$ and $\hat{C}_j$. The false positive is a sum of number of individuals that are in  $\hat{C}_j$ but not in $C_i$. and the false negative is a sum of number of individuals that are in $C_i$  but not in $\hat{C}_j$. 
 

%% file: 06-results.tex
\section{Results}
\subsection{Leadership dynamics}
\begin{figure}[!ht]
\begin{center}
\includegraphics[width=.8\columnwidth]{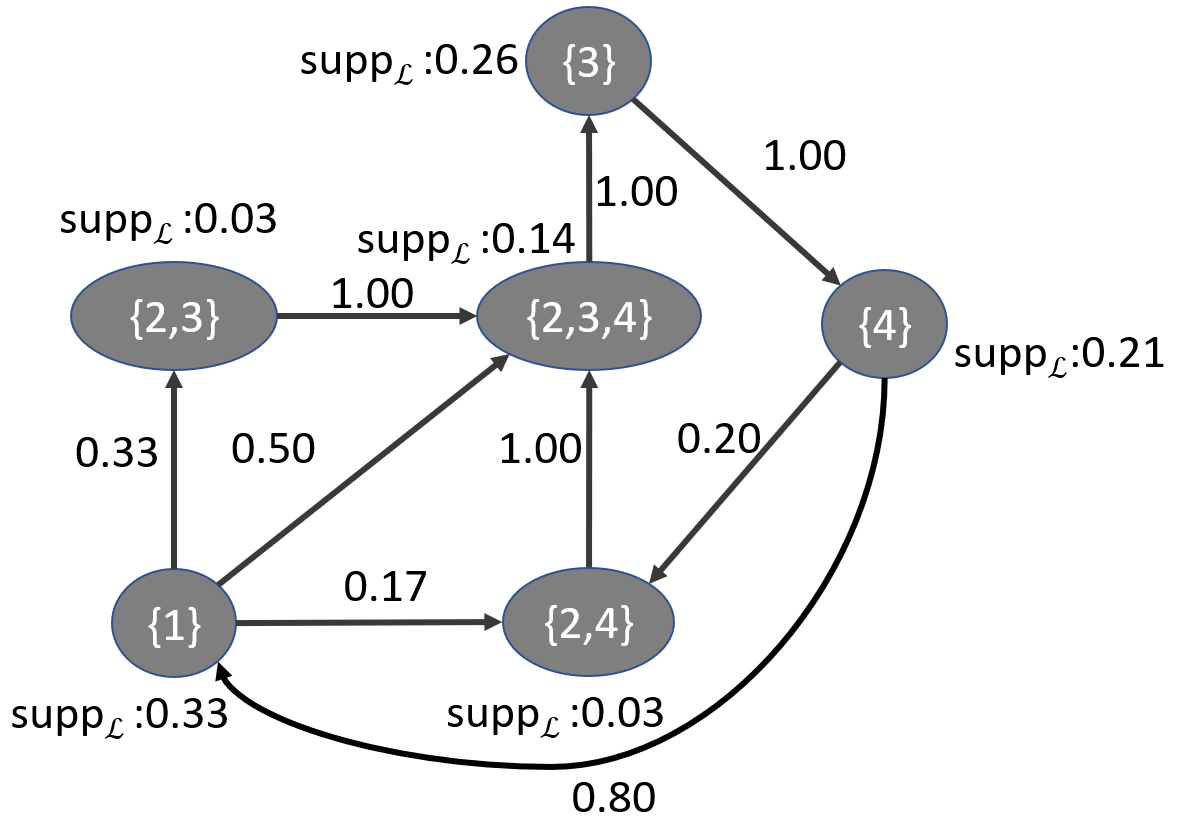}
\end{center}
\caption{The example of the inferred diagram of leadership dynamics by our framework from a Type-1-HM dataset. Comparing the inferred diagram with the ground truth, only nodes $\{2,4\}$ and $\{2,3\}$ are false positive nodes. The support of $\{1\},\{2,3,4\},\{3\}$ and $\{4\}$ should be $0.25$, and our framework can infer the support for each node closely to $0.25$. } 
\label{fig:HMType1StateMachineEx}
\end{figure}

\begin{figure}[!ht]
\begin{center}
\includegraphics[width=.5\columnwidth]{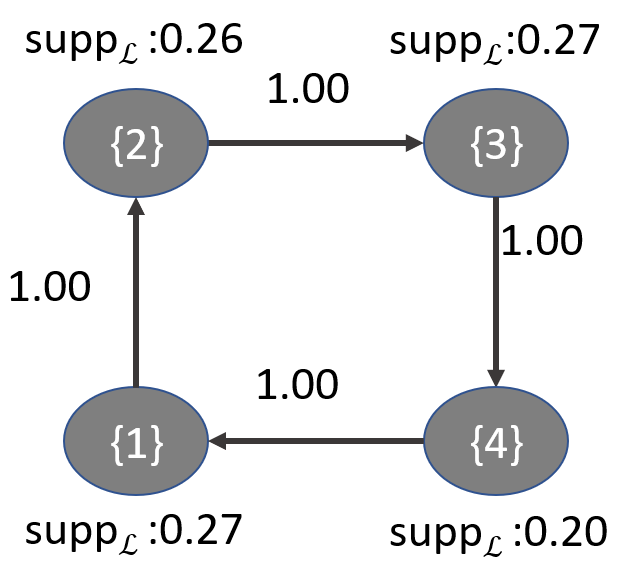}
\end{center}
\caption{The example of the inferred diagram of leadership dynamics by our framework using a Type-2-HM dataset. Comparing the inferred diagram with the ground truth, there are no false positive nodes. The support of $\{1\},\{2\},\{3\}$ and $\{4\}$ should be $0.25$, and our framework can infer the support for each node closely to $0.25$}
\label{fig:HMType2StateMachineEx}
\end{figure}

 We set the time window parameter $\omega$ using the inference method in~\cite{mFLICASDM18}. Fig.~\ref{fig:HMType1StateMachineEx} and Fig.~\ref{fig:HMType2StateMachineEx} show the examples of  inferred diagrams of leadership dynamics by our framework from Type-1-HM (Hierarchical model with Splitting/Merging coordination events) and Type-2-HM (Hierarchical model with Linear coordination events) datasets respectively. In Fig.~\ref{fig:HMType1StateMachineEx}, comparing the inferred diagram with the ground truth, only nodes $\{2,4\}$ and $\{2,3\}$ are false positive nodes, both with very low support of 0.03. 
 
 This implies that despite the complex dynamics of leadership in Type-1-Dynamics case, our framework was still able to retrieve the diagram of leadership dynamics accurately. For the Type-2-HM dataset, which is less complex than Type1-HM case, Fig.~\ref{fig:HMType2StateMachineEx} shows that there are no false positive nodes in the inferred diagram. Moreover, in both Type-1-HM and Type-2-HM cases, the support of each node should be $0.25$, and our framework can infer the support for each node closely to $0.25$.\\
 
 Regarding the mining sequence patterns of leadership dynamics described in Section~\ref{sec:Miningmethod}, Table~\ref{table:suppPathmining} shows an example of max-support sequences of leadership dynamics that our framework reported from HM datasets. In both dynamics types, the sequences are consistent with the ground truth in Fig.~\ref{fig:Dytype}. 
 
\begin{table}[]
\centering
\caption{The example of sequences of leadership dynamics that have the highest support from HM datasets.}
\label{table:suppPathmining}
\begin{tabular}{|c|c|c|}
\hline
\textbf{Datasets} & \multicolumn{1}{c|}{\textbf{Sequences}} & \multicolumn{1}{l|}{$\mathrm{supp}_{path}(\mathcal{L},P_{i,j})$} \\ \hline
\textbf{Type-1-HM} & \{2,3,4\},\{3\},\{4\},\{1\}     & 0.71                                  \\ \hline
\textbf{Type-2-HM} & \{1\},\{2\},\{3\},\{4\}         & 0.95                                  \\ \hline
\end{tabular}
\end{table}

 Next, we compared our framework, which uses the following networks concept~\cite{mFLICASDM18}, to the method based on direction networks proposed in FLOCK method~\cite{andersson2008reporting} to infer a diagram of leadership dynamics. In direction networks, at any time $t$, if $i$ is moving toward the same direction as $j$ but $j$ is in front of $i$, then $i$ follows $j$. The median of all loss distributions in both  Type-1 and Type-2 dynamics datasets are reported in Table~\ref{table:lossTable}.  The first row of Table~\ref{table:lossTable} shows the distribution of loss values (Eq.~\ref{eq:lossFunc}) in Type-1-dynamics datasets. The direction network approach was reasonably competitive for the Type-1 dynamics. We were able to use the direction networks to infer the states with splits and merges but the change of leadership was often missed by this underlying method. Not surprisingly, then, the direction network-based method performed significantly worse than the following network-based approach for the Type-2 dynamics. Qualitatively, and as a distribution of the loss values overall,  the following networks as the basis for the diagram inference performed better than the direction networks in our framework. In the second row of Table~\ref{table:lossTable}, the following networks also performed better than direction networks in Type-2-dynamics datasets.



\begin{table}[]
\centering
\caption{The median of loss values in the prediction task of diagrams of leadership dynamics.}
\label{table:lossTable}
\begin{tabular}{|p{.7in}|c|c|c|c|c|c|}
\hline
\textbf{Dyn. Type}     & \multicolumn{3}{c|}{\textbf{Type 1}}    & \multicolumn{3}{c|}{\textbf{Type 2}}    \\ \hline
\textbf{Model}            & \textbf{HM} & \textbf{DM} & \textbf{IC} & \textbf{HM} & \textbf{DM} & \textbf{IC} \\ \hline
\textbf{Following Network} & 0.13       & 0.19           & 0.24        & 0           & 0.03        & 0.08        \\ \hline
\textbf{Direction Network} & 0.19            & 0.19            & 0.25           & 0.19            & 0.19            & 0.25           \\ \hline
\end{tabular}
\end{table}

 In Table~\ref{table:Htestres}, we reported the hypothesis testing results of the significance of leadership-event order (Section~\ref{sec:sig1}).  With respect to the type of the leadership model, for the HM, which is a well-structure model, the inferred diagrams are more significantly different from the null-model diagram than for the other leadership models. With respect to the types of the dynamics, in the complex type-1-dynamics datasets  our framework inferred diagrams that are more significantly different from the null model.  Lastly, the following networks were able to infer diagrams that are more different from the null model than the direction networks.
 
\begin{table}[]
\centering
\caption{Hypothesis testing results of the significance of leadership-event order in Section~\ref{sec:sig1}. We reject $H_0$ at $\alpha=0.01$. Each element in the table represents the percentage of the times when the tests successfully reject $H_0$. }
\label{table:Htestres}
\begin{tabular}{|p{.7in}|c|c|c|c|c|c|}
\hline
\textbf{Dyn. Type}     & \multicolumn{3}{c|}{\textbf{Type 1}}    & \multicolumn{3}{c|}{\textbf{Type 2}}    \\ \hline
\textbf{Model}            & \textbf{HM} & \textbf{DM} & \textbf{IC} & \textbf{HM} & \textbf{DM} & \textbf{IC} \\ \hline
\textbf{Following Network} & 0.99        & 0.55        & 0.38        & 0.86        & 0.08        & 0.20        \\ \hline
\textbf{Direction Network} & 0.00        & 0.00        & 0.06        & 0.00        & 0.00        & 0.06        \\ \hline
\end{tabular}
\end{table}

For hypothesis testing of the significance of frequencies of leadership-event sequences (Section~\ref{sec:sig2}), the result is shown in Table~\ref{table:HPathtestres}. Similar to the the edge-weight distribution testing, the support distributions of the well-structure model, HM, are significantly different from the support distribution of the null model. The following networks also can be used to infer diagrams that are different from the rewiring diagrams than the direction networks based approach. However, in the simple type-2-dynamics datasets, our framework was able to infer diagrams that are more different from the null model compared to the complex type-1-dynamics case.   

\begin{table}[]
\centering
\caption{Hypothesis testing results of the significance of frequencies of leadership-event sequences in Section~\ref{sec:sig2}.  We reject $H_0$ at $\alpha=0.01$. Each element in the table represents the percentage of the case when the test successfully rejects $H_0$.}
\label{table:HPathtestres}
\begin{tabular}{|p{.7in}|c|c|c|c|c|c|}
\hline
\textbf{Dyn. Type}     & \multicolumn{3}{c|}{\textbf{Type 1}}    & \multicolumn{3}{c|}{\textbf{Type 2}}    \\ \hline
\textbf{Model}            & \textbf{HM} & \textbf{DM} & \textbf{IC} & \textbf{HM} & \textbf{DM} & \textbf{IC} \\ \hline
\textbf{Following Network} & 0.95           & 0.35        & 0.23        & 0.94        & 0.84        & 0.66           \\ \hline
\textbf{Direction Network} & 0.07        & 0.08        & 0.20        & 0.07        & 0.07        & 0.20           \\ \hline
\end{tabular}
\end{table}

\begin{figure}[!ht]
\includegraphics[width=1\columnwidth]{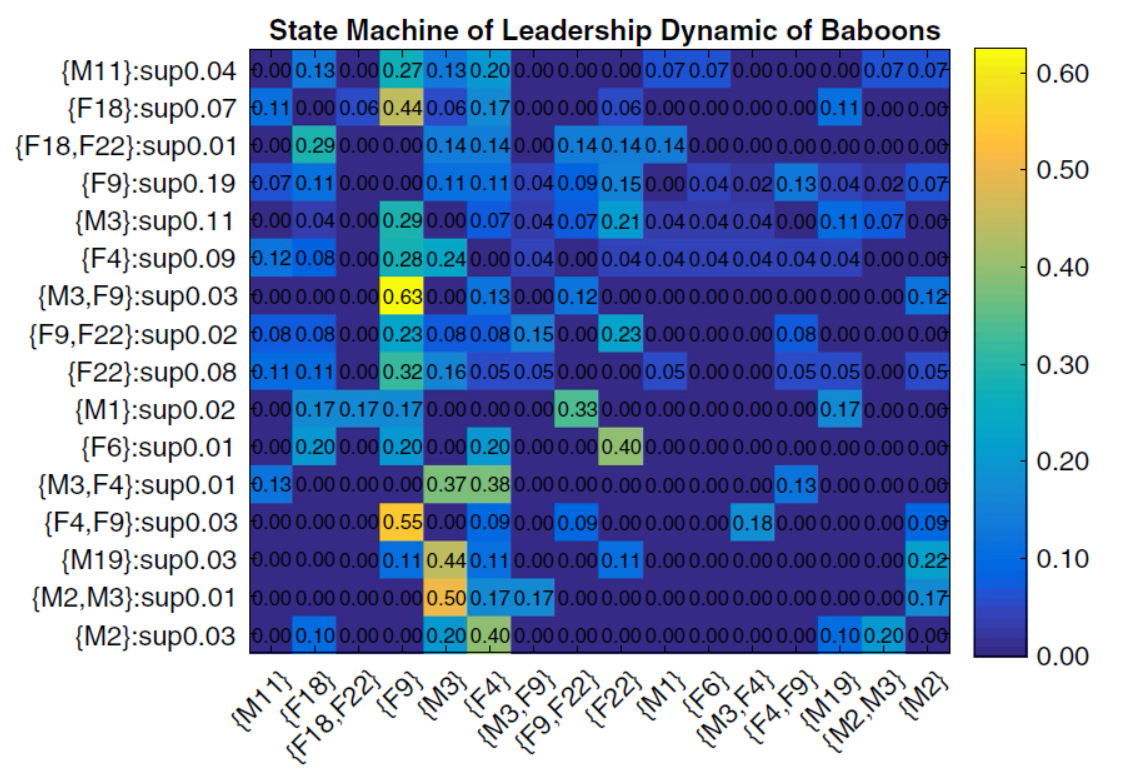}
\caption{The inferred diagram of leadership dynamics of the baboon dataset from our framework. Each row represents the node of leader sets of previous state and each column represents the next state. Each row label consists of baboon gender: `M' or `F', a set of frequent-leader IDs, and the support value of frequent-leader set. For example, in row 3 and column 2, the event that two female baboons F18 and F22 are leading their separate sub-groups concurrently can happen with the support 0.1 (out of all the coordination times). These two sub-groups have a chance to be merged together to a larger group lead by F18 with the probability 0.29.  }
\label{fig:BaboonRes}
\end{figure}

For the baboon dataset, we reported the information that we can retrieve from the dataset using our framework as a case study. Fig.~\ref{fig:BaboonRes} shows the inferred diagram of leadership dynamics from our framework. Each row represents the node of leader sets of previous state and each column represents the next state. Each row label consists of baboon gender: `M' or `F', a set of frequent-leader IDs, and the support value of frequent-leader set. For example, in row 3 and column 2, the event that two female baboons F18 and F22 are leading their separate sub-groups concurrently can happen with the support 0.1 (out of all the coordination times). These two sub-groups have a chance to be merged together to a larger group lead by F18 with the probability 0.29. In 4th column ($\{$F9$\}$), we found that no matter what the previous sub-groups were, there was a high chance that the next group would be lead solely  by the female baboon F9. In 4th row, F9 has the highest support (0.19), which means F9 (who happens to be the dominant female) often leads the troop, with the next highest support of 0.11 for the male baboon M3 (5th column, the alpha male). Lastly, at row  5 and column 4, if M3 and F9 are leading their separate sub-groups, then the two groups will be merged to a larger group lead by F9 with  probability 0.63. 

The hypothesis testing of the edge-weight distribution shows that the baboon's diagram is significantly different from the null model, with  100\% of the time the tests successfully rejecting $H_0$. However, for the hypothesis testing of sequence-support distributions, the baboons' sequences of  leadership dynamics are not significantly different from the rewired diagram.  Only 5\% of the times the tests successfully reject $H_0$. This indicates that while individual leaders identity is non-random and pairwise leadership transition patterns are significant, there are no leadership sequences that often appear significantly within the baboon dataset. Nevertheless, Table~\ref{table:baboonPath} shows  baboons' sequences of  leadership dynamics that have the top-4 highest support. This result is the evidence that F9 is an important individual who frequently leads the group.

\begin{table}[]
\centering
\caption{Baboons' sequences of  leadership dynamics that have the top-4 highest support}
\label{table:baboonPath}
\begin{tabular}{|c|c|c|}
\hline
\textbf{Baboon} & \multicolumn{1}{c|}{\textbf{Sequences}} & \multicolumn{1}{l|}{$\mathrm{supp}_{path}(\mathcal{L},P_{i,j})$} \\ \hline
\textbf{Seq. 1} & \{M11\},\{F9\},\{M3\}             & 0.0354                                \\ \hline
\textbf{Seq. 2} & \{M18\},\{F9\},\{M3\}             & 0.0354                                \\ \hline
\textbf{Seq. 3} & \{M18\},\{F9\},\{M22\}            & 0.0354                                \\ \hline
\textbf{Seq. 4} & \{M4\},\{F9\},\{M2\}              & 0.0354                                \\ \hline
\end{tabular}
\end{table}

These results show that our framework provides the opportunity for scientists to gain more insight into their datasets in order to generate scientific hypotheses, which might lead to important scientific discoveries (in this case, about the collective behavior and leadership dynamics of social animals).

\subsection{Followership dynamics}
\subsubsection{Co-faction and lead-follow networks}
\begin{figure}[!ht]
\includegraphics[width=1\textwidth]{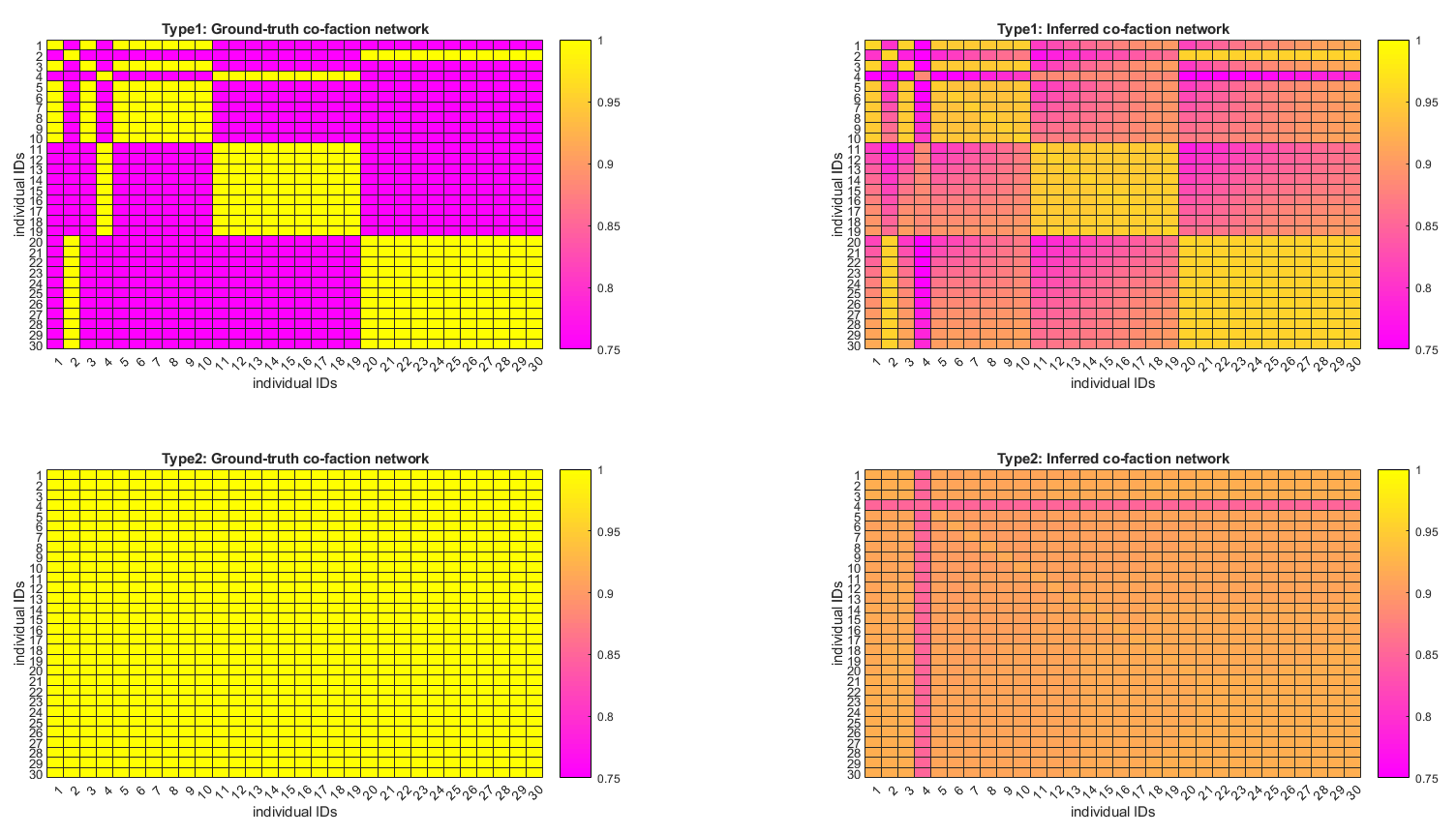}
\caption{The adjacency matrices of ground-truth and predicted co-faction networks. (Top-left) the ground-truth matrix of Type-1 dynamics. (Top-right) the predicted matrix of Type-1 dynamics. (Bottom-left) the ground-truth matrix of Type-2 dynamics. (Bottom-right) the predicted matrix of Type-2 dynamics. Each predicted matrix is the result of aggregation of  co-faction adjacency matrices from 100 datasets. The lighter color implies a higher value of $\mathrm{csupp}_{\mathcal{F}}(i,j)$. }
\label{fig:cofactComp} 
\end{figure}



Fig.~\ref{fig:cofactComp} shows the results of ground-truth and predicted adjacency matrices of co-faction network by our framework from Type-1-HM (Hierarchical model with Splitting/Merging coordination events) and Type-2-HM (Hierarchical model with Linear coordination events) datasets. Each predicted matrix is the result of aggregation of  co-faction adjacency matrices from 100 datasets. The result shows that our inferred matrices are mostly similar to the ground-truth matrices with some variation due to noise. ID4 has the highest error in Fig.~\ref{fig:cofactComp} since it appears during the interval when the group stop moving. Because mFLICA is designed to handle movement initiation analysis, it has a limitation to analyze stopping intervals of movement. Hence, mFLICA cannot capture the behavior of a leader ID4 well.  

\begin{figure}[!ht]
\includegraphics[width=1\textwidth]{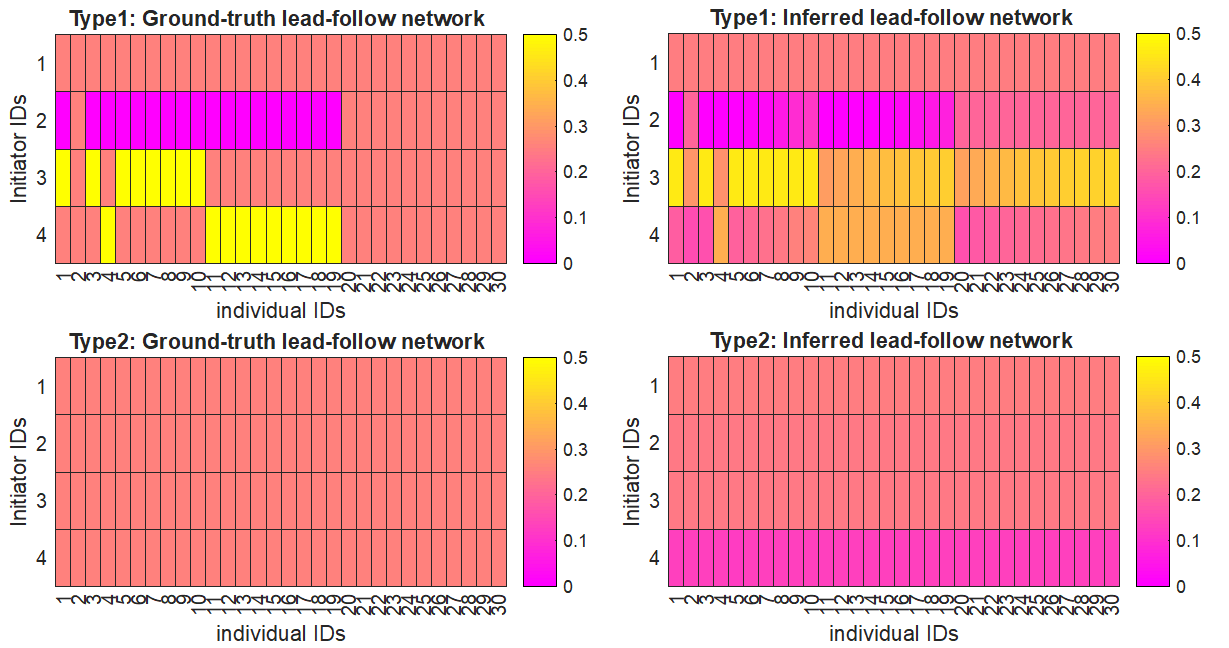}
\caption{The adjacency matrices of ground-truth and predicted lead-follow networks. (Top-left) the ground-truth matrix of Type-1 dynamics. (Top-right) the predicted matrix of Type-1 dynamics. (Bottom-left) the ground-truth matrix of Type-2 dynamics. (Bottom-right) the predicted matrix of Type-2 dynamics. Each predicted matrix is the result of aggregation of  lead-follow adjacency matrices from 100 datasets. The lighter color implies a higher value of $\mathrm{lfsupp}_{\mathcal{F}}(i,L)$ where $i$ is a column individual (follower) and $L$ is a row individual (initiator). }
\label{fig:leadfollComp} 
\end{figure}

Fig.~\ref{fig:leadfollComp} shows the results of ground-truth and predicted adjacency matrices of lead-follow networks with $\phi_{LF}=0.1$. Each predicted matrix is the result of aggregation of  lead-follow adjacency matrices from 100 datasets. The result also shows that our inferred matrices are mostly similar to the ground-truth matrices with some variation. ID4 result has the highest error because of mFLICA limitation that we have just discussed.

\begin{table}[]
\centering
\caption{Loss values of co-faction and lead-follow networks inference. Each element represents a mean of loss value $\pm$ two standard deviations from 100 datasets. A lower value implies a better performance of inference. }
\label{table:cofactAndleadfollLoss}
\begin{small}
\begin{tabular}{l|c|c|c|c|}
\cline{2-5}
 & \multicolumn{2}{c|}{Following Network} & \multicolumn{2}{c|}{Direction Network} \\ \cline{2-5} 
 & Type-1  & Type-2  & Type-1  & Type-2  \\ \hline
\multicolumn{1}{|l|}{Co-fact loss} & $0.184\pm 0.013$ & $0.187\pm 0.030$ & $0.398\pm 0.014$ & $0.451\pm 0.011$ \\ \hline
\multicolumn{1}{|l|}{Lead-foll loss} & $0.054\pm 0.012$ & $0.026\pm 0.001$ & $0.063\pm 0.007$ & $0.057\pm0.002$ \\ \hline
\end{tabular}
\end{small}
\end{table}

We also report the quantitative result of prediction of both co-faction and lead-follow networks using following networks compared with direction networks in Table~\ref{table:cofactAndleadfollLoss}.  Overall, our proposed framework using following networks performed better than the direction network framework. For co-faction networks, the loss values are higher than lead-follow network loss values. This implies that finding who are in the same faction frequently is a bit harder than finding who are loyal members of specific leaders.  

\subsubsection{Clustering results}

\begin{table}[]
\centering
\caption{Q-value in Eq.~\ref{eq:Qeq} of clustering results. Each element represents a mean of Q-value value $\pm$ two standard deviations from 100 datasets. We expect Type-1 dynamics to have higher $Q$-value since there are three strong clusters, while  Type-2 dynamics should have the $Q-value$ around zero. We report the results from our framework and the candidate approach NM~\cite{PhysRevE.69.066133}. }
\label{table:Qtable}
\begin{tabular}{l|l|l|}
\cline{2-3}
 & Our framework & NM community detection \\ \hline
\multicolumn{1}{|l|}{Type-1 dynamics} & $0.6934\pm0$ & $0.460\pm 0.145$ \\ \hline
\multicolumn{1}{|l|}{Type-2 dynamics} & $0.064\pm0$ & $0.066\pm0$ \\ \hline
\end{tabular}
\end{table}

In  the clustering task, given a co-faction network as an input, we compared our proposed framework with a standard community detection algorithm in~\cite{PhysRevE.69.066133}. The NM community detection method greedily searches for the partition of individuals that maximize the Q-value in Eq.~\ref{eq:Qeq}. Table~\ref{table:Qtable} shows the result of Q-values of our framework and NM community detection. For type-1 dynamics, we should have a high value of $Q$-value since there are three strong clusters. Table~\ref{table:Qtable} shows that even though NM method tried to find the best clustering partition that maximizes Q-value, our framework found a set of better clusters that has a higher Q-value than NM's clusters. For type-2 dynamics, since there is only one cluster, we expect that the Q-value should be close to zero. Both methods performed well in this case.

\begin{table}[]
\centering
\caption{F1-score of ground-truth vs. inferred clustering results. Each element represents a mean of F1-score value $\pm$ two standard deviations from 100 datasets. A higher F1 score value implies a better performance of clustering inference. We report the results from our framework and the candidate approach NM~\cite{PhysRevE.69.066133}. }
\label{table:F1clusttable}
\begin{tabular}{l|l|l|}
\cline{2-3}
 & Our framework & NM clustering \\ \hline
\multicolumn{1}{|l|}{Type-1 dynamics} & $0.983 \pm 0.003$ & $0.940\pm 0.096$ \\ \hline
\multicolumn{1}{|l|}{Type-2 dynamics} & $1\pm 0$ & $0.983 \pm 0$ \\ \hline
\end{tabular}
\end{table}

We also reported the results of clustering comparison between the ground-truth and inferred clusters. The result in Table~\ref{table:F1clusttable} shows that our framework performed better than NM in both types of dynamics. 

\subsection{Baboon followership dynamics}
\begin{figure}[!ht]
\includegraphics[width=.9\textwidth]{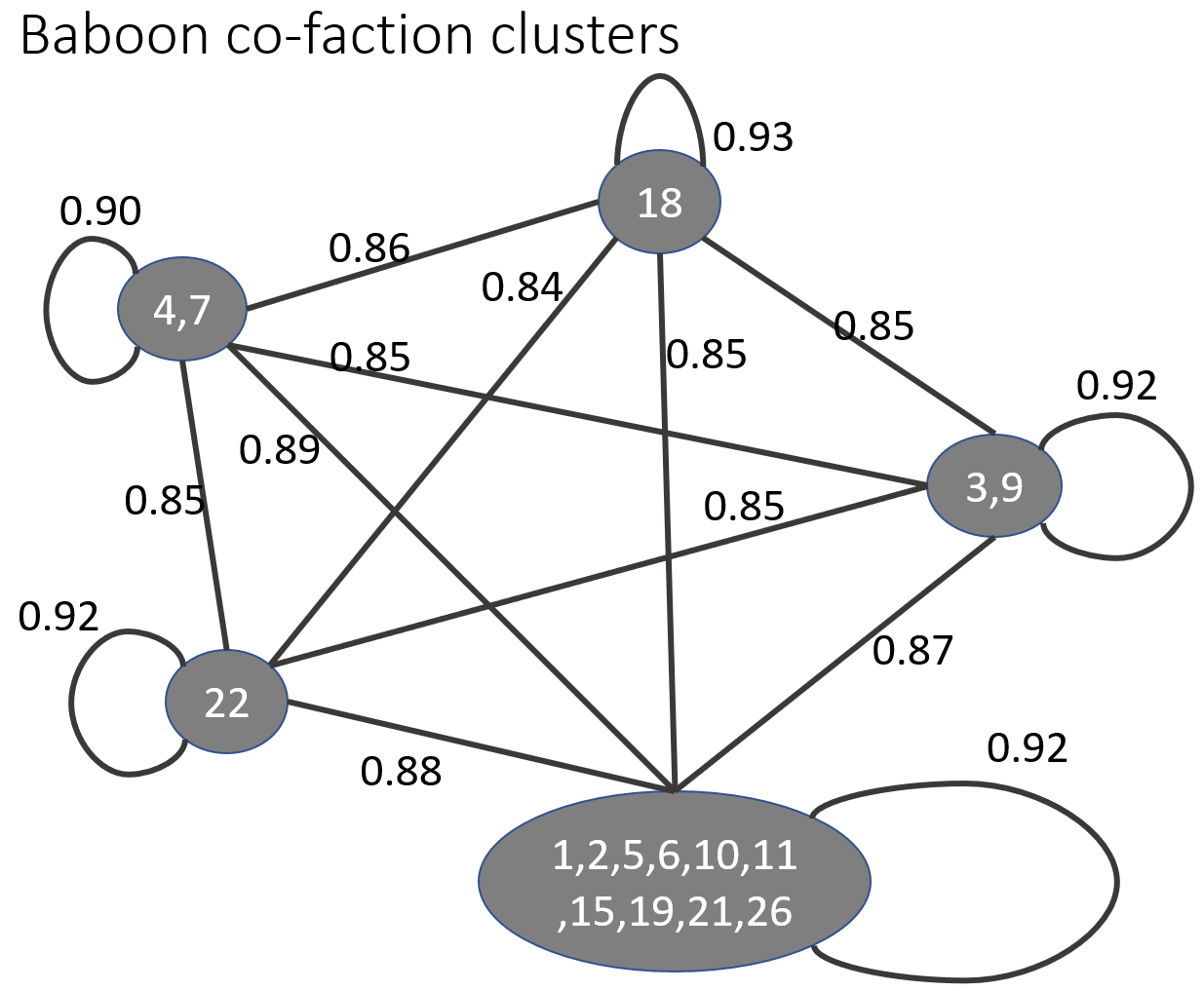}
\caption{The co-faction clusters of the baboons dataset inferred by our framework. Each node is a cluster labeled with IDs of cluster members and each edge is a median of $\mathrm{csupp}_{\mathcal{F}}$ of members between clusters. }
\label{fig:BBcofactclusters} 
\end{figure}

\begin{figure}[!ht]
\includegraphics[width=1\textwidth]{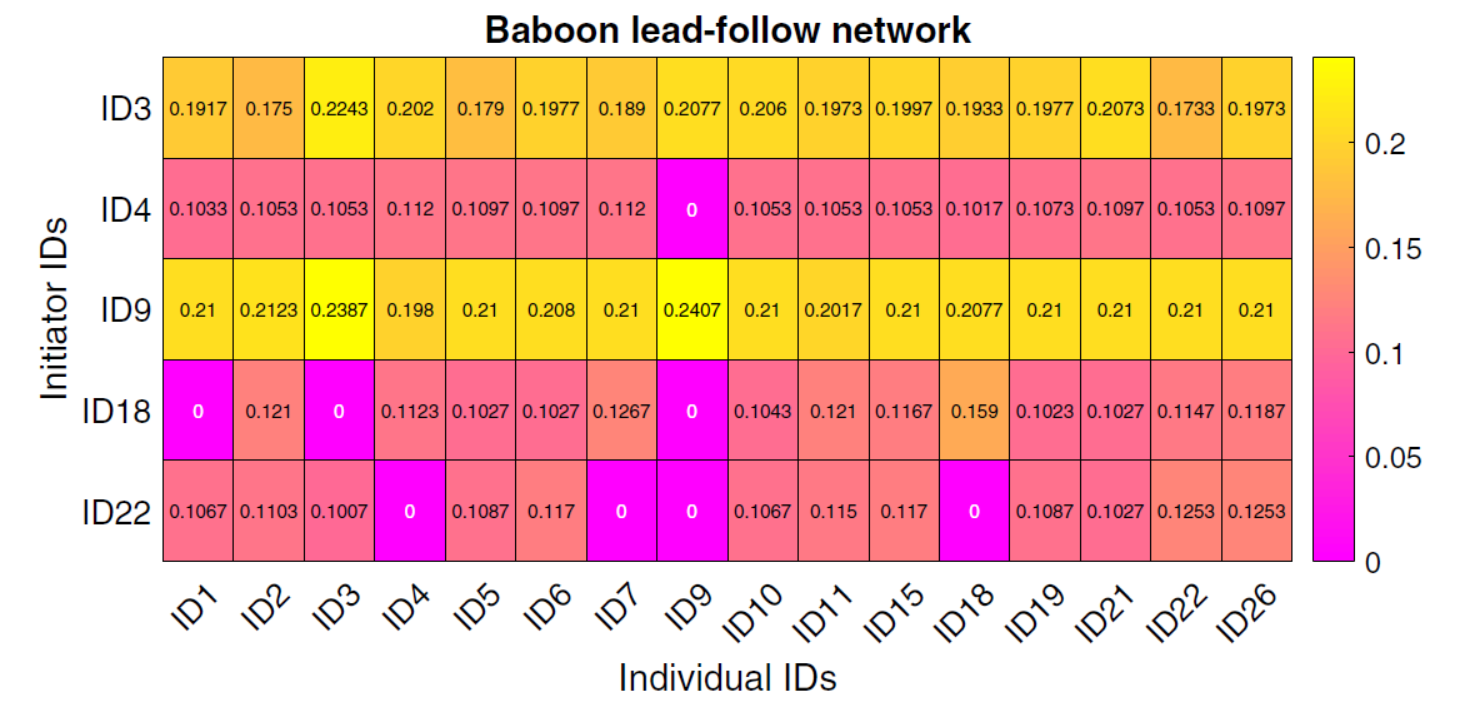}
\caption{The lead-follow network of the baboon dataset inferred by our framework.}
\label{fig:BBleadfoll} 
\end{figure}

For the baboon dataset, we reported the result of co-faction clustering (Fig.~\ref{fig:BBcofactclusters}) and lead-follow network (Fig.~\ref{fig:BBleadfoll}) inferred from the trajectories of baboon during pre-coordination intervals of high-coordination events.  

Fig.~\ref{fig:BBcofactclusters} shows five major clusters in the dataset. The interesting cluster is the cluster of ID3 and ID9. ID3 is an alpha male while ID9 is an alpha female. Since they are in the same cluster, this implies that they might be a couple. 

Fig.~\ref{fig:BBcofactclusters} shows a lead-follow network of the troop. It shows that ID3 and ID9 frequently act as initiators of the group that everyone follows. In both Fig.~\ref{fig:BBcofactclusters} and \ref{fig:BBleadfoll} support the hypothesis that ID3 and ID9 might be a center of influence of the group decision making.

%% file: 07-conclusion.tex
\section{Conclusion}
In this paper, we proposed a new approach to analyze time series of group movement data. We formalized a new computational problem, \slip, and \sfip, as well as proposed a framework as a solution of these problems. Our framework can be used to address several questions regarding leadership and followership dynamics of group movement, such as `what is the probability of having two sub-groups lead by $i$ and $j$ being merged together to be a larger group lead by $k$ later?', `what is the frequency of having $i$ and $k$ co-leading their sub-groups concurrently?', `how likely is it that a specific sub-group that $i$ is a member will be leading by an individual $j$ from the same faction?', etc. We use the leadership inference framework, mFLICA~\cite{mFLICASDM18}, to infer the time series of leaders and their factions from movement datasets, then propose an approach to mine and model frequent patterns of both leadership and followership dynamics. We evaluate our framework performance by using several simulated datasets, as well as  the real-world dataset of  baboon movement to demonstrate the applications of our framework.  These are novel computational  problems and, to the best of our knowledge, there are no existing comparable methods to address them. Thus, we modify and extend an existing leadership inference framework to provide a non-trivial baseline for comparison. Our framework performs better than this baseline in all datasets. Our framework opens the opportunities for scientists to generate testable scientific hypotheses about the dynamics of leadership in movement data.